\documentclass[letterpaper,12pt]{article}
\usepackage{color,amssymb,enumerate,float,amsmath}

\usepackage{ifpdf}
\ifpdf
	\usepackage[pdftex]{graphicx}
	\usepackage[pdftex,unicode,implicit]{hyperref}

	\hypersetup{
  	pdftitle     = {}, 
  	pdfkeywords  = {},
  	pdfauthor    = {},
  	pdfcreator   = {pdf\LaTeXe\ with package \flqq hyperref\frqq},
  	pdfproducer  = {pdf\LaTeXe\ with package \flqq hyperref\frqq},
  	pdfpagemode  = UseNone,  
  	pdffitwindow = true,  
  	unicode      = true,
  	plainpages   = true,
  	colorlinks   = true,  
  	citecolor    = black,  
  	urlcolor     = blue, 
  	linkcolor    = black
	}

\else

  \usepackage[dvips]{graphicx}

\fi 

\makeatletter
\@addtoreset{equation}{section}
\makeatother

\begin{document}
	
\thispagestyle{empty}

\begin{center}
{\bf \LARGE Quantum Lagrangian of the Ho\v{r}ava theory and its nonlocalities}
\vspace*{15mm}

{\large Jorge Bellor\'{\i}n}$^{1}$,
{\large Claudio B\'orquez}$^{2}$
{\large and Byron Droguett}$^{3}$
\vspace{3ex}

{\it Department of Physics, Universidad de Antofagasta, 1240000 Antofagasta, Chile.}
\vspace{3ex}

$^1${\tt jorge.bellorin@uantof.cl,} \hspace{.5em}
$^2${\tt cl.borquezg@gmail.com} \hspace{.5em}
$^3${\tt byron.droguett@ua.cl}

\vspace*{15mm}
{\bf Abstract}
\begin{quotation}{\small\noindent
We perform the BFV quantization of the $2+1$ projectable and the $3+1$ nonprojectable versions of the Ho\v{r}ava theory. This is a Hamiltonian formalism, and noncanonical gauges can be used with it. In the projectable case, we show that the integration on canonical momenta reproduces the quantum Lagrangian known from the proof of renormalization of Barvinsky et al. This quantum Lagrangian is nonlocal, its nonlocality originally arose as a consequence of getting regular propagators. The matching of the BFV quantization with the quantum Lagrangian reinforces the program of quantization of the Ho\v{r}ava theory. We introduce a local gauge-fixing condition, hence a local Hamiltonian, that leads to the nonlocality of the Lagrangian after the integration. For the case of the nonprojectable theory, this procedure allows us to obtain the complete (nonlocal) quantum Lagrangian that takes into account the second-class contraints. We compare with the integration in general relativity, making clear the relationship between the underlying anisotropic symmetry of the Ho\v{r}ava theory and the nonlocality of its quantum Lagrangian. 
 }
\end{quotation}
\end{center}

\thispagestyle{empty}

\newpage
\section{Introduction}
Several studies have been devoted to the consistent quantization of the Ho\v{r}ava theory \cite{Horava:2009uw}. Some of the analyses performed under the framework of quantum field theory can be found in Refs.~\cite{Barvinsky:2015kil,Barvinsky:2017kob,Contillo:2013fua,D'Odorico:2014iha,D'Odorico:2015yaa,Orlando:2009en,Pospelov:2010mp,Colombo:2014lta,Benedetti:2013pya,Griffin:2017wvh,Barvinsky:2019rwn,Bellorin:2019gsc,Barvinsky:2021ubv}. Other approaches of quantization, as causal dynamical triangulations and loop quantum gravity has been done, for example in Refs.~\cite{Ambjorn:2010hu,Anderson:2011bj,Ambjorn:2013joa,Li:2014bla,Zhang:2020smo}. A fundamental advance is the renormalizability proof of the projectable version presented in Ref.~\cite{Barvinsky:2015kil}. The difference between the projectable and the nonprojectable versions of the Ho\v{r}ava theory is that in the former the lapse function is restricted to be a function only on time, a condition that can be imposed consistently in the Ho\v{r}ava theory, whereas in the latter it can be a general function of time and space. An interesting feature of the proof of renormalizability is the introduction of nonlocal gauge-fixing conditions, which leads to a nonlocal quantum Lagrangian. The nonlocal gauges were motivated by the goal of obtaining regular propagators for all quantum modes, such that the renormalizability can be achieved in a similar way to the case of Lorentz-violating gauge theories \cite{Anselmi:2007ri,Anselmi:2008bq,Anselmi:2008bs}. The condition of regularity implies that the propagators have no divergences in space valid for each time and viceversa. For the case of the Ho\v{r}ava theory, the propagators acquire anisotropic higher order in momentum space.

Due to the emphasis on the symmetry, quantization of gauge field theories are usually performed in the Lagrangian formalism, rather than in the Hamiltonian formalism. The standard procedure for fixing the gauge is the Faddeev-Popov method \cite{Faddeev:1967fc}, together with its associated Becchi-Rouet-Stora-Tyutin (BRST) symmetry \cite{Becchi:1975nq}. Nevertheless, the quantization of the Ho\v{r}ava theory using the Hamiltonian formalism deserves to be considered. In particular, the quantization of the nonprojectable case is a delicate issue since it is a theory with second-class constraints. The analogous of the Hamiltonian constraint of general relativity acquires a second-class behavior in the nonprojectable Ho\v{r}ava theory, which can be related to the reduction of the gauge symmetry. The Hamiltonian formalism provides a natural framework for the quantization of theories with second-class constraints. Indeed, the contribution to the measure of these constraints is defined in the phase space \cite{Senjanovic:1976br}. Analyses on the Hamiltonian formulation and the dynamics of the degrees of freedom of the Ho\v{r}ava theory can be found in Refs.~\cite{Kobakhidze:2009zr,Bellorin:2010je,Loll:2014xja,Kluson:2010nf,Donnelly:2011df,Bellorin:2011ff}.

The nonlocal gauge-fixing conditions introduced in the projectable case are noncanonical gauges, in the sense that they involve a Lagrange multiplier. If one wants to use this kind of gauges in the Hamiltonian formalism, then an extension of the phase space is required. Motivated by this, two of us presented the Batalin-Fradkin-Vilkovisky (BFV) quantization of the $2+1$ nonprojectable Ho\v{r}ava theory in Ref.~\cite{Bellorin:2021tkk}. The BFV formalism provides a quite general framework for quantization of systems with constraints, with the particularity that first-class constraints are not imposed explicitly and their Lagrange multipliers are promoted to be part of the canonical variables. The BFV formalism was first presented in Ref.~\cite{Fradkin:1975cq} as a way to introduce noncanonical gauge-fixing conditions in the Hamiltonian formalism. This extension allows us to introduce relativistic gauges in the phase space, which is a way to establish the unitarity of relativistic gauge theories. The formalism was extended in Ref.~\cite{Batalin:1977pb} to the case of systems with fermionic constraints, together with establishing an essential connection with the BRST symmetry. In Ref.~\cite{Fradkin:1977xi}, the BFV quantization was extended to the case of systems with (bosonic and fermionic) second-class constraints, and for the general case when the Hamiltonian and the BRST charge have expansions of certain order on the ghosts fields. In Ref.~\cite{Fradkin:1975sj} the formalism was applied to general relativity. In the general BFV scheme of quantization, the gauge symmetry is fixed by the choice of a fermionic function, and the resulting gauge-fixed path integral enjoys a BRST symmetry operating on the canonical variables, as we commented.

Since the quantization of the Ho\v{r}ava theory has been focused in the Lagrangian and the Hamiltonian BFV formalisms, a natural question is to ask whether both approaches are equivalent. Indeed, in Ref.~\cite{Barvinsky:2015kil} it is indicated that the introduction of an extra variable, that can be interpreted as the conjugate momentum of the shift vector, eliminates the nonlocality of the final Lagrangian. This suggests that we perform a complete Hamiltonian analysis of the quantum theory. In this paper we undertake this problem, with interest in connecting with the gauge-fixing conditions known from the renormalization of the projectable case \cite{Barvinsky:2015kil}. We study the BFV quantization of the $2+1$ projectable and $3+1$ nonprojectable versions of the Ho\v{r}ava theory. We emphasize that this provides a consistent quantization in the sense that it is based on a canonical phase space that admits the gauges of interest. This is particularly critical for the nonprojectable case. After we present the BFV quantization, we get the quantum Lagrangian for each case by means of integration on the canonical momenta, including the ghosts. To achieve this, we specialize the form of the gauge-fixing condition, by specifying its functional dependence on the momenta (it depends on a particular momentum). This leads us to introduce operators that, after the integration, yield the nonlocalities on the quantum Lagrangian. In the case of the projectable theory we arrive at the same quantum Lagrangian defined in Ref.~\cite{Barvinsky:2015kil}. In the nonprojectable case we obtain the quantum Lagrangian of the theory where the second-class constraints have been taken into account. In this way the quantum Lagrangian of the nonprojectable case is consistent. We compare the same method of integration in general relativity, whose BFV quantization was presented in Ref.~\cite{Fradkin:1975sj}, noticing that the relativity leads to a local quantum Lagrangian.

This paper is organized as follows. In section 2 we present the BFV quantization of the projectable theory and the integration to obtain the quantum Lagrangian. We first develop the formalism for a Ho\v{r}ava theory defined in a general spatial dimension $d$, and eventually we specialize to the $2+1$ case. In section 3 we present the same approach for the $3+1$ nonprojectable case, presenting the BFV quantization and the quantum Lagrangian. In section 4 we compare with general relativity. Finally, we present some conclusions. 


\section{Projectable Ho\v{r}ava theory}
\subsection{Classical theory}
The Ho\v{r}ava theory \cite{Horava:2009uw}, both in the projectable and nonprojectable cases, is based on a given foliation that has an absolute physical meaning. The aim is to get an anisotropic scaling at the ultraviolet that favor the renormalizability of the theory, where a parameter $z$ measures the degree of anisotropy. To hold this anisotropic scaling, the dimensions of the space and time are defined to be
\begin{equation}
	[t]=-z,\qquad [x^i]=-1 \,.
\end{equation}
The order $z$ is fixed by the criterium of power-counting renormalizability, which yields $z = d$, where $d$ is the spatial dimension of the foliation. The Arnowitt-Deser-Misner variables $N$, $N^i$ and $g_{ij}$ are used to describe the gravitational dynamics on the foliation. The allowed coordinate transformations on the foliation,
\begin{equation}
	\delta t=f(t),\qquad \delta x^{i}=\zeta^{i}(t,\vec{x}) \,,
\label{fdiffcoord}
\end{equation}
lead to the gauge symmetry of the foliation-preserving diffeomorphisms,
\begin{eqnarray}
	\delta N &=& 
	\zeta^{k} \partial_{k} N + f \dot{N} + \dot{f}N \,, 
	\label{deltaN}
	\\ 
	\delta N_{i} &=& 
	\zeta^{k} \partial_{k} N_{i} + N_{k} \partial_{i} \zeta^{k} + \dot{\zeta}^{j} g_{ij} + f \dot{N}_{i} + \dot{f}N_{i} \,,
	\\
	\delta g_{ij} &=& \zeta^{k} \partial_{k} g_{ij} + 2g_{k(i}\partial_{j)} \zeta^{k} + f \dot{g}_{ij} 
\end{eqnarray}
(strictly, the spatial diffeomorphisms are the gauge transformations). The condition that defines the projectable version is that the lapse function is restricted to be a function only of time, $N = N(t)$, a condition that is preserved by the transformation (\ref{deltaN}). In this section we summarize the canonical formulation of the projectable case, dealing with an arbitrary number $d$ of spatial dimensions. The Hamiltonian analysis of the projectable case, taking the infrared effective action, was done in Ref.~\cite{Kobakhidze:2009zr}. Further analyses, with different boundary conditions, can be found in Ref.~\cite{Loll:2014xja}. The quantization of the same model under the scheme of loop quantum gravity has been studied in Ref.~\cite{Zhang:2020smo}.

The Lagrangian of the projectable theory is given by
\begin{equation}
	\mathcal{L} =
	\sqrt{g} N \Big( K^{ij}K_{ij}-\lambda K^{2}-\mathcal{V}[g_{ij}] \Big) \,,
\label{lagrangian}
\end{equation}
where the extrinsic curvature is defined by
\begin{equation}
	K_{ij} =
	\frac{1}{2N}\Big(\dot{g}_{ij}-2\nabla_{(i}N_{j)}\Big)
\label{extrinsiccurvature}
\end{equation}
and $\mathcal{V}[g_{ij}]$, called the potential, is built from invariants of the spatial curvature and their derivatives, up to the order $2z$.

In the Hamiltonian formulation the canonical pair is $(g_{ij}\,,\pi^{ij})$, whereas $N(t)$ and $N^i(t,\vec{x})$ enter as Lagrange multipliers. Since $N(t)$ is a function only of time, there is an associated global constraint, given in terms of a spatial integral. This constraint is
\begin{equation}
	\int d^dx \mathcal{H} = 0 \,,
	\quad
	\mathcal{H} \equiv
	\frac{1}{\sqrt{g}}\Big(\pi^{ij}\pi_{ij}+\frac{\lambda}{1-d\lambda}\pi^{2}\Big)+\sqrt{g}\mathcal{V} \,.
\label{globalconstraint}
\end{equation} 
Throughout this paper we assume that $\lambda$ does not take the critical value $\lambda = 1/d$. This global constraint does not eliminate a complete functional degree of freedom. The local constraint of the theory is the momentum constraint,
\begin{equation}
\mathcal{H}_{i} = -2\nabla^k\pi_{ki} \,.
\end{equation}
The primary Hamiltonian is
\begin{equation}
	{H}_0 = \int d^dx \mathcal{H}_0 =
	N \int d^dx \mathcal{H} \,.
	\label{primaryhamiltonianprojectable}
\end{equation}
Since $N$ is a function of time in the projectable theory, we take advantage of the symmetry of reparameterizing the time, Eqs.~(\ref{fdiffcoord}) and (\ref{deltaN}), to set $N = 1$. With this setting the primary Hamiltonian density is equivalent to $\mathcal{H}$. Due to their importance in the BFV quantization, and since the Hamiltonian is equivalent to $\mathcal{H}$, we show the following two brackets between constraints,
\begin{eqnarray}
	&&
	\left\{\int d^{d}x\epsilon^{k}\mathcal{H}_{k} \,, \int d^{d}y\eta^{l}\mathcal{H}_{l}\right\} = 
	\int d^{d}x\mathcal{H}_{l}\mathcal{L}_{\vec{\epsilon}}\eta^{l},
	\label{diffalgebra}
	\\ &&
	\left\{\int d^{d}x\epsilon^{k}\mathcal{H}_{k} \,, \rho \int d^{d}y\mathcal{H}\right\} = 0 \,.
	\label{hih0}
\end{eqnarray}
In the above $\rho$ is a test function only of time whereas $\epsilon^k$ and $\eta^k$ are test functions of time and space.

\subsection{BFV quantization}
The initial consideration in the BFV formalism is that the constrained system under quantization must be involutive. This means that, given a Hamiltonian $\mathcal{H}_{0}$ and a set of functions $G_{a}$, the following relations are satisfied
\begin{align}
	&\left\lbrace G_{a},G_{b}\right\rbrace = U^{c}_{ab}G_{c},
	\label{algebraBFV}
	\\
	&\left\lbrace\mathcal{H}_{0},G_{a}\right\rbrace
	=V^{b}_{a}G_{b}.
	\label{bracketH0}
\end{align}
To avoid writing huge expressions, we use a simplification on the notation of brackets: we insert densities instead of spatial integrals, such as $\{ A \,, B \} \rightarrow \{ \int d^d x A \,, \int d^dy B \}$. 
The first-class constraints are part of the definition of the $G_a$ functions. The other part is given by the canonical momenta conjugated to the Lagrange multipliers of the first-class constraints, since these multipliers are promoted to canonical variables in the BFV extension of the phase space. The extended phase space is completed with the canonical pair of fermionic ghosts $(\eta^{a},\mathcal{P}_{a})$, where each pair is incorporated for each function $G_a$.

To apply this formalism to the projectable Ho\v{r}ava theory, we identify the momentum constraint $\mathcal{H}_i$ as the only first-class constraint, being the shift vector $N^i$ its Lagrange multiplier. By denote by $\pi_i$ the canonical momentum conjugated to $N^i$. Thus, the functions are $G_a = ( \mathcal{H}_i \,, \pi_i )$. Since $\pi_i$ commutes with itself and with $\mathcal{H}_i$, the algebra (\ref{algebraBFV}) reduces to the algebra of $\mathcal{H}_i$,
\begin{equation}
 \{ \mathcal{H}_i \,, \mathcal{H}_j \} =
 U_{ij}^k \mathcal{H}_k \,.
\end{equation}
This corresponds to the algebra of spatial diffeomorphisms, as shown in (\ref{diffalgebra}), and we take the definition of $U_{ij}^k$ from it. $U_{ab}^c = 0$ for $a,b,c > i$. The primary Hamiltonian is identified in (\ref{primaryhamiltonianprojectable}), hence the bracket (\ref{bracketH0}) corresponds to (\ref{hih0}), such that $V_a^b = 0$. By incorporating the ghost fields, the full BFV phase space of the projectable Ho\v{r}ava theory is given by the canonical pairs $(g_{ij},\pi^{ij})$, $(N_{i},\pi^{i})$ and $(\eta_{a},\mathcal{P}^{a})$.  The ghosts can be split in the two sets, $(\eta_1^i , \mathcal{P}^1_i )$, $(\eta_2^i , \mathcal{P}^2_i )$. The gauge-fixing condition is incorporated in the path integral by means of a fermionic function $\Psi$, which is a given functional on the extended phase space. Thus, the BFV path integral of the projectable Ho\v rava theory is given by
\begin{equation}
 Z=
 \int \mathcal{D} g_{ij} \mathcal{D}\pi^{ij} \mathcal{D}N^{k} \mathcal{D}\pi_{k} \mathcal{D}\eta^{a} \mathcal{D}\mathcal{P}_{a}
 \exp\left[ i \int dt d^{d}x 
 \Big(\pi^{ij}\dot{g_{ij}} +\pi_{k}\dot{N}^{k}+\mathcal{P}_{a}\dot{\eta}^{a}-\mathcal{H}_{\Psi}\Big) \right]
 \,.
\end{equation}
In this formalism the ghosts eliminate the unphysical quantum degrees of freedom that should be eliminated by the first-class constraints. Indeed, in $d$ spatial dimensions the canonical pairs $(g_{ij},\pi^{ij})$, $(N_{i},\pi^{i})$ amount for $d(d+3)$ degrees of freedom, and the ghosts $(\eta_1^i , \mathcal{P}^1_i )$, $(\eta_2^i , \mathcal{P}^2_i )$ sum $4d$ degrees. After subtracting, one gets $d(d-1)$ physical degrees of freedom in the phase space of the quantum theory. In $d=2$ this yield $2$ degrees of freedom, which represent the scalar mode of the $2+1$ projectable theory in canonical variables. In $d=3$ the degrees of freedom are six, which are the two tensorial modes plus the extra scalar mode. Since the Ho\v{r}ava theory has anisotropic scaling, it is important to write down the dimensions of the several fields. This is 
\begin{equation}
\begin{array}{lll}
 & [g_{ij}]=0 \,, \qquad  & [\pi^{ij}]=d=z \,, \\[1ex]
 & [N^k]=z-1 \,, \qquad & [\pi_k]=1+d-z = 1 \,, \\[1ex]
 & [\eta_1^i] = [\mathcal{P}^2_i] = (d-z)/2 = 0 \,, \qquad
 & [\eta_2^i] = [\mathcal{P}^1_i] = (d+z)/2 = z \,.
\end{array}
\label{dimensions}	
\end{equation}

In the general BFV formalism, the gauge-fixed quantum Hamiltonian is defined by
\begin{equation}
	\mathcal{H}_{\Psi}=\mathcal{H}_{1}+\left\lbrace\Psi,\Omega\right\rbrace \,.
\label{generalgfhamiltonian}
\end{equation}
The Poisson bracket is extended to include fermionic variables,
\begin{equation}
 \{ A , B \} =
 \dfrac{\delta^{\mbox{\tiny R}} A}{\delta q^r} 
 \dfrac{\delta^{\mbox{\tiny L}} B}{\delta p^r} -
 (-1)^{n_A n_B} \dfrac{\delta^{\mbox{\tiny R}} B}{\delta q^r} \dfrac{\delta^{\mbox{\tiny L}} A}{\delta p^r} \,, 
\end{equation} 
where R and L denote right and left derivatives and $n_A$ is 0 or 1 depending on whether $A$ is a boson or a fermion. $\Omega$ is the generator of the BRST symmetry. According to the extension of the BFV formalism presented in Ref.~\cite{Fradkin:1977xi}, $\Omega$ and $\mathcal{H}_{1}$ are defined in terms of expansions in the ghost fields,
\begin{align}
	\Omega&=G_{a}\eta^{a}+\sum_{k=1}^{s}\mathcal{P}_{b_{k}}\ldots\mathcal{P}_{b_{1}}\Omega^{b_{1}\ldots b_{k}} \,,
	\label{omega}
	\\
	\mathcal{H}_{1}&=\mathcal{H}_{0}+\sum_{k=1}^{s}\mathcal{P}_{b_{k}}\ldots\mathcal{P}_{b_{1}}\mathcal{H}_{1}^{b_{1}\ldots b_{k}},
	\label{H1}
\end{align}
where $s$ represents the rank of theory. The coefficient functions of the first order in $\mathcal{P}_{a}$ are given by
\begin{equation}
	\Omega^{a}=-\frac{1}{2}U^{a}_{bc}\eta^{b}\eta^{c} \,,
	\qquad
	\mathcal{H}_{1}^{a}=V^{a}_{b}\eta^{b} \,.
\end{equation}
The rest of coefficients, up to the order $s$ of the theory, are obtained by recurrence relations, starting from the first-order ones \cite{Fradkin:1977xi}. An essential condition of the BFV formalism is that $\Omega$ and $\mathcal{H}_1$ must satisfy
\begin{equation}
	\{ \Omega \,, \Omega\} = 0 \,,
	\qquad 
	\{ \mathcal{H}_{1} \,, \Omega \} = 0 \,.
\label{bfvconditions}	
\end{equation}
The first one is a nontrivial condition since $\Omega$ is a fermionic variables. These conditions support the BRST symmetry of the quantum theory.

The projectable Ho\v{r}ava theory is of first order, that is, $\Omega$ ends at the first order in the ghosts, whereas $\mathcal{H}_1$ is of zeroth order,
\begin{eqnarray}
	&&
	\Omega = 
	G_{a} \eta^{a} - \frac{1}{2} U^{c}_{ab} \eta^{a} \eta^{b}  \mathcal{P}_{c}
	= 
	\mathcal{H}_{k} \eta_{1}^{k} + \pi_{k} \eta_{2}^{k} 
	- \frac{1}{2} U^{k}_{ij} \eta_{1}^{i} \eta_{1}^{j} \mathcal{P}_{k}^{1} \,,
	\\ &&
	\mathcal{H}_{1} = \mathcal{H}_{0} = \mathcal{H} \,.
\end{eqnarray} 
The conditions (\ref{bfvconditions}) are satisfied as follows. We have the bracket of $\Omega$ with itself, 
\begin{equation}
	\{\Omega \,, \Omega\} = 
	\{ \mathcal{H}_i \eta^i_1 \,, \mathcal{H}_j \eta^j_1 \}
	- \{ \mathcal{H}_i \eta^i_1 \,,  
	U^{k}_{mn} \eta_{1}^{m} \eta_{1}^{n} \mathcal{P}_{k}^{1} \}
	+ \frac{1}{4} \{ U^{k}_{ij} \eta_{1}^{i} \eta_{1}^{j} \mathcal{P}_{k}^{1} \,, 
	U^{l}_{mn} \eta_{1}^{m} \eta_{1}^{n} \mathcal{P}_{l}^{1} \} \,.
	\label{OMEGAOMEGAcap4}
\end{equation}
The first two brackets are equal,
\begin{equation}
	\{ \mathcal{H}_i \eta^i_1 \,, \mathcal{H}_j \eta^j_1 \} = 
	\{ \mathcal{H}_i \eta^i_1 \,,  
	U^{k}_{mn} \eta_{1}^{m} \eta_{1}^{n} \mathcal{P}_{k}^{1} \} = 
	U^i_{jk} \eta_{1}^{j} \eta_{1}^{k} \mathcal{H}_i \,,
\end{equation}
hence cancel themselves. The last bracket is proportional to the structure $\eta_1^j \eta_1^m \eta_1^n$ $U^k_{ij} U^i_{mn}$, which is zero by the Jacobi identity. Therefore $\{ \Omega \,, \Omega \} = 0$. Next,
\begin{equation}
	\{ \mathcal{H}_{1} \,, \Omega\} = 
	\{ \mathcal{H} \,, \Omega \} =
	\{ \mathcal{H} \,, \eta_{1}^i \mathcal{H}_i \} = 0 \,,
\end{equation}
where the last equality follows from (\ref{hih0})\footnote{Recalling that we mean spatial integrals inside the brackets.}. Therefore, we obtain the BFV gauge-fixed Hamiltonian of the projectable Ho\v{r}ava theory,
\begin{equation}
	\mathcal{H}_{\Psi} =
	\mathcal{H}
	+ \left\{ \Psi \,, \mathcal{H}_{k} \eta_{1}^{k} + \pi_{k} \eta_{2}^{k} 
	- \frac{1}{2} U^{k}_{ij} \eta_{1}^{i} \eta_{1}^{j} \mathcal{P}_{k}^{1} \right\} \,.
\label{gaugefixingcap4}
\end{equation}

According to the original BFV formulation, $\Psi$ can adopt a form suitable for relativistic gauges. It turns out that this form is also suitable for the anisotropic symmetry of the Ho\v{r}ava theory. First, we deal with gauge-fixing conditions of the general structure
\begin{equation}\label{gaugefixing}
	\Phi^{i}=-\dot{N}^{i}+\chi^{i} = 0 \,,
\end{equation}
where the phase-space functional $\chi^{i}$ is the part of the gauge-fixing condition that can be chosen. Thus, the specific BFV fermionic gauge-fixing function is
\begin{equation}\label{Psi}
	\Psi=
	\mathcal{P}_{i}^{1}N^{i}+\mathcal{P}_{i}^{2}\chi^{i} \,.
\end{equation}
With this choice the gauge-fixed Hamiltonian becomes
\begin{equation}
 \mathcal{H}_{\Psi} = 
 \mathcal{H} 
 +\mathcal{H}_kN^{k}
 + \mathcal{P}_k^{1} \eta_2^{k}
 - \mathcal{P}^{1}_{i} \left(N^{j}\partial_{j}\eta_{1}^{i} + N^{i}\partial_{j}\eta_{1}^{j}\right)
 +\{\mathcal{P}_i^{2}\chi^{i} \,, \Omega \} \,.
\end{equation}
Throughout this paper we assume that the gauge-fixing condition $\chi^l$ does not depend on the ghosts fields, then the Hamiltonian takes the form
\begin{equation}
	\mathcal{H}_{\Psi} = 
	\mathcal{H} 
	+\mathcal{H}_kN^{k}
	+ \mathcal{P}_k^{1} \eta_2^{k}
	- \mathcal{P}^{1}_{i} \left(N^{j}\partial_{j}\eta_{1}^{i} + N^{i}\partial_{j}\eta_{1}^{j}\right)
	+ \pi_k\chi^{k}
	+ \mathcal{P}_i^{2}\{\chi^{i} \,, \mathcal{H}_k\} \eta_1^{k}
	+ \mathcal{P}_i^{2} \dfrac{ \delta\chi^{i}}{\delta N^{l}} \eta_2^{l} \,.
\end{equation}
Therefore, the BFV path integral for the projectable Ho\v{r}ava theory in the gauge (\ref{gaugefixing}) -- (\ref{Psi}) becomes
\begin{equation}
\begin{split}
	Z =&
	\int \mathcal{D} g_{ij} \mathcal{D}\pi^{ij} \mathcal{D}N^{k} \mathcal{D}\pi_{k} \mathcal{D}\eta_1^i \mathcal{D}\mathcal{P}^1_i \mathcal{D}\eta_2^i \mathcal{D}\mathcal{P}^2_i
    \\ & 
	\exp\left[ i \int dt d^{d}x 
	\Big(\pi^{ij}\dot{g_{ij}} +\pi_{k}\dot{N}^{k}
	+\mathcal{P}^1_i \dot{\eta}_1^i + \mathcal{P}^2_i \dot{\eta}_2^i
	- \mathcal{H} 
	- \mathcal{H}_kN^{k}
	\right. 
    \\ & \left. 
	- \mathcal{P}_k^{1} \eta_2^{k}
    + \mathcal{P}^{1}_{i}\left(N^{j}\partial_{j}\eta_{1}^{i} + N^{i}\partial_{j}\eta_{1}^{j}\right) 
	- \pi_k\chi^{k}
    - \mathcal{P}_i^{2}\{\chi^{i} \,, \mathcal{H}_k\} \eta_1^{k}
    - \mathcal{P}_i^{2} \dfrac{ \delta\chi^{i}}{\delta N^{l}} \eta_2^{l}
	\Big) \right]
	\,.
\end{split}
\label{bfvfinalprojectable}
\end{equation}

The generator of the BRST symmetry $\Omega$ acts on the canonical fields by means of the canonical transformation
\begin{equation}
	\tilde{\varphi}=\varphi+\left\lbrace\varphi,\Omega\right\rbrace \epsilon \,,
\end{equation}
where $\epsilon$ is the fermionic parameter of the transformation. The transformation of the fields is
\begin{equation}
\begin{array}{lll}
	&
	\delta_{\Omega}g_{ij}=
	2g_{k(i}\nabla_{j)}\eta^{k}_{1}\epsilon \,,
	\qquad\qquad
	&
	\delta_{\Omega}\pi^{ij}=
	-2\pi^{k(i}\nabla_{k}\eta^{j)}_{1}\epsilon
	   +\nabla_{k}(\pi^{ij}\eta^{k}_{1})\epsilon \,,
	\\[1ex] \nonumber
	&\delta_{\Omega}N^{k}=
	\eta^{k}_{2}\epsilon \,,
	\qquad\qquad\qquad\qquad
	&
	\delta_{\Omega}\pi_{k}=0 \,,
	\\[1ex] \nonumber
	&\delta_{\Omega}\eta^{i}_{1}=
	-\frac{1}{2}U^{i}_{jk}\eta^{j}_{1}\eta^{k}_{1}\epsilon \,,
	\qquad\qquad\;\;\;\;
	&
	\delta_{\Omega}\mathcal{P}^{1}_{i}=
	\mathcal{H}_{i}\epsilon
	   -U^{k}_{ij}\eta^{j}_{1}\mathcal{P}^{1}_{k}\epsilon \,,
	\\[1ex] \nonumber
	&\delta_{\Omega}\eta^{k}_{2}=0 \,,
	\qquad\qquad\qquad\qquad\qquad
	&
	\delta_{\Omega}\mathcal{P}^{2}_{k}=\pi_{k}\epsilon \,.
\end{array}
\end{equation}

\subsection{Quantum Lagrangian}
We continue with working on an arbitrary spatial dimensionality $d$, eventually we specialize to the $d=2$ case. For the BFV quantization we have defined the structure of the gauge-fixing condition (\ref{gaugefixing}), which has the part $\chi^i$ unspecified. To arrive at the quantum Lagrangian, we impose conditions on the functional form of $\chi^i$ that allow us to perform the integration on the several canonical momenta. These conditions allow us to make a connection with the same gauge fixing used in the proof of renormalizability of the projectable theory.

We start with the integration on the momentum $\pi_i$. The term $-\pi_k \chi^k$ in the action of (\ref{bfvfinalprojectable}) suggests to demand that $\chi^i$ has a linear dependence on $\pi_i$, leading to a quadratic term in $\pi_i$ in the Hamiltonian, otherwise a higher order dependence on this variable could lead to a violation of unitarity, which is contradiction to the spirit of the Ho\v{r}ava theory and its anisotropic symmetry. Therefore, we assume the structure of the gauge-fixing condition  
\begin{equation}
	\chi^{k}= 
	\mathfrak{D}^{ki}\pi_{i} + \Gamma^{k}[g_{ij},N^{k}] \,,
\label{gaugepi}
\end{equation}
where $\Gamma^k$ is a functional that may depend only on $g_{ij}$ and $N^k$. The restriction that $\Gamma^k$ does not depend on the momentum $\pi^{ij}$ allows us to perform the integration straightforwardly. According to the anisotropic dimensional assignments (\ref{dimensions}), the gauge-fixing condition must satisfy $[\chi^k]=2z-1 $, hence the dimension of the operator $\mathfrak{D}^{ij}$ must be 
\begin{equation}
[\mathfrak{D}^{ij}]=3z-d-2 = 2z - 2 \,. 
\end{equation}
Below we give explicitly the operator $\mathfrak{D}^{ij}$ and the gauge-fixing form $\Gamma^i$ in the perturbative framework. Nevertheless, many operations can be carried out without recurring to perturbations and for general $\Gamma^i$. Hence we stay for a while on nonperturbative variables, using only the fact that $\mathfrak{D}^{ij}$ is a flat operator (does not depend on any field variable).

By setting the form (\ref{gaugepi}) for the gauge-fixing condition, the last three terms of the action of Eq.~(\ref{bfvfinalprojectable}) become
\begin{equation}
	- \pi_k \mathfrak{D}^{ki} \pi_{i}
	- \pi_k\Gamma^{k}
	- \mathcal{P}_i^{2}\{ \Gamma^{i} \,, \mathcal{H}_k\}\eta_1^{k}
	- \mathcal{P}_i^{2}\dfrac{\delta\Gamma^{i}}{\delta N^{l}}\eta_2^{l} 
	\,.
\end{equation}
We may complete the square involving $\pi_i$ and then integrate on the shifted variable, obtaining the path integral
\begin{equation}
\begin{split}
	Z = &
	\int \mathcal{D} g_{ij} \mathcal{D}\pi^{ij}\mathcal{D} N^{k}\mathcal{D}\eta^{a}\mathcal{D}\mathcal{P}_{a} 
    \\ &
	\exp\left[ i \int dt d^{d}x 
	\Big( \pi^{ij}\dot{g_{ij}} + \mathcal{P}_{a}\dot{\eta}^{a}
	+ \frac{1}{4} (\dot{N}^{k}-\Gamma^{k}) \mathfrak{D}^{-1}_{kl} (\dot{N}^{l}-\Gamma^{l})
	- \mathcal{H}
	- \mathcal{H}_k N^{k} 
	\right.  \\ &
	\left.
	- \mathcal{P}_k^{1}\eta_2^{k}
	+ \mathcal{P}^{1}_{i}\left(N^{j}\partial_{j}\eta_{1}^{i} + N^{i}\partial_{j}\eta_{1}^{j}\right)	
	- \mathcal{P}_i^{2} \{ \Gamma^{i} \,, \mathcal{H}_k \} \eta_1^{k}
	- \mathcal{P}_i^{2} \dfrac{\delta\Gamma^{i}}{\delta N^{l}} \eta_2^{l} 
	\Big) \right] \,.
\end{split}
\label{actionintegratedpi}
\end{equation}
Since $\mathfrak{D}^{ij}$ is a local operator, its inverse $\mathfrak{D}_{kl}^{-1}$, which has arisen by the integration, is a nonlocal operator.

Now we move to the ghost sector. The following change of notation is useful for the final the quantum Lagrangian:
\begin{equation}
\begin{split}
\eta_1^i \rightarrow C^i \,, \qquad & 
   \mathcal{P}^1_i \rightarrow \bar{\mathcal{P}}_i \,,
\\
\eta_2^i \rightarrow \mathcal{P}^i \,, \qquad & 
   \mathcal{P}^2_i \rightarrow \bar{C}_i \,.
\end{split}
\label{newnotation}
\end{equation}
We may perform the integration of the Grassmann variables $\mathcal{P}^i$ and $\bar{\mathcal{P}}_i$, which arise in the action (\ref{actionintegratedpi}) in the terms
\begin{equation}
	- \bar{\mathcal{P}}_{k}\mathcal{P}^{k} + \bar{\mathcal{P}}_{k}\left(\dot{C}^{k}-N^{j}\partial_{j}C^{k}-N^{k}\partial_{j}C^{j}\right) + \mathcal{P}^{k}\left(\dot{\bar{C}}_k+\bar{C}_i\frac{\delta\Gamma^{i}}{\delta N^{k}}\right)  \;.
	\label{squaregrassmann}
\end{equation}
The bilinear $-\bar{\mathcal{P}}_k \mathcal{P}^k$ can be completed, such that the Gaussian integration on these Grassmann variables can be performed (without consequences on the measure). After these steps of integration, the path integral becomes
\begin{equation}
	\begin{split}
		Z = &
		\int \mathcal{D} g_{ij} \mathcal{D}\pi^{ij}\mathcal{D} N^{k}\mathcal{D}\bar{C}_i \mathcal{D} C^i
		\\ &
		\exp\left[ i \int dt d^{d}x 
		\Big( \pi^{ij}\dot{g_{ij}} 
		+ \frac{1}{4} (\dot{N}^{k}-\Gamma^{k}) \mathfrak{D}^{-1}_{kl} (\dot{N}^{l}-\Gamma^{l})
		- \mathcal{H} - \mathcal{H}_k N^{k}
		\right. \\
		& \left.	 
		+\left(\dot{C}^{k}-N^{j}\partial_{j}C^{k}-N^{k}\partial_{j}C^{j}\right) 
		\left( \dot{\bar{C}}_k 
		+ \bar{C}_i \frac{ \delta\Gamma^{i}}{\delta N^k} \right)				
		- \bar{C}_i \{ \Gamma^{i} \,, \mathcal{H}_k \} C^{k} 
		\Big) \right] \,.
	\end{split}
	\label{prepiij}
\end{equation}

Now we focus the integration on $\pi^{ij}$. A significant part of the computations can be continued on nonperturbative grounds. Since this is interesting on its own, in appendix \ref{integratepi} we show this nonperturbative integration for the case of the projectable theory. In what follows we adopt a perturbative approach. We consider perturbations around the analogous of the Minkowski spacetime, given by $g_{ij} = \delta_{ij}$, $\pi^{ij} = 0$, $N^i = 0$, $\bar{C}_i = C^i = 0$.

We comment that for the $d=2$-dimensional case we take the operator $\mathfrak{D}^{ij}$ as 
\begin{equation}
	\mathfrak{D}^{ij} = \delta_{ij} \Delta + \kappa \partial_{i}\partial_{j}  \,,
	\quad
	\mathfrak{D}^{-1}_{ij} =
	\frac{\delta_{ij}}{\Delta} 
	- \frac{\kappa}{1+\kappa} \frac{\partial_i \partial_j }{\Delta^2} \,,
	\label{D}
\end{equation}
where $\kappa$ is an arbitrary constant. The inverse $\mathfrak{D}^{-1}_{ij}$ is a nonlocal operator of dimension $-2$ in $d=2$. The operator $\mathfrak{D}_{ij}^{-1}$ (\ref{D}) was introduced in the gauge-fixing condition used in Ref.~\cite{Barvinsky:2015kil}, with the aim of introducing the nonlocality that finally leads to regular propagators. This version of the operator $\mathfrak{D}^{ij}$ for the $d=2$ case arises in several steps of the integration for arbitrary dimension $d$, with a fixed value of $\kappa$. For this reason we denote these special cases as 
\begin{align}
 & \mathfrak{D}^{ij}_1 =  \delta_{ij} \Delta + \partial_{i}\partial_{j} \,,
 \label{D1} \\
 & \mathfrak{D}_2^{ij}=
   \delta_{ij}\Delta
   + \frac{1+\lambda}{1-\lambda}\partial_{i}\partial_{j} \,,
 \label{hatD} \\
 & \mathfrak{D}^{ij}_3 = 
   \delta_{ij} \Delta + (1-2\lambda)\partial_i\partial_j \,.
\end{align}
The inverse of $\mathfrak{D}_2^{ij}$ is also required,
\begin{equation}
\mathfrak{D}_{2ij}^{-1} =
\frac{\delta_{ij}}{\Delta}
- \frac{1+\lambda}{2} \frac{\partial_{i}\partial_{j}}{\Delta^{2}} \,.
\end{equation}
Note that the operator $\mathfrak{D}_2^{ij}$ cannot be extended to the relativistic limit $\lambda = 1$. 

We denote the perturbative variables
\begin{equation}
 g_{ij} - \delta_{ij} =  h_{ij} \,,
 \qquad
 \pi^{ij} = p^{ij} \,,
 \qquad
 N^i = n^i \,,
\label{perturbations}
\end{equation}
and the ghosts $\bar{C}_i$, $C^i$ are considered perturbative variables of first order. The quantum action given in (\ref{prepiij}), expanded up to quadratic order, results
\begin{equation}
	\begin{split}
		S =&
		\int dt d^{d}x 
		\left\lbrace p^{ij}\dot{h}_{ij} 
		+\frac{1}{4} (\dot{n}^{k}-\Gamma^{k}) \mathfrak{D}^{-1}_{kl} (\dot{n}^{l}-\Gamma^{l}) -\mathcal{H} - n^{k} \mathcal{H}_k
		\right.\\
		&\left.
		+\dot{C}^{k} \left( \dot{\bar{C}}_{k} 
		+\bar{C}_i \frac{\delta\Gamma^{i}}{\delta n^{k}} \right) -\bar{C}_i\{\Gamma^{i},\mathcal{H}_k\} C^{k}
		\right\rbrace \,,
	\end{split}
\label{actionpiijproj}
\end{equation}
where
\begin{eqnarray}
	\mathcal{H}_{j} &=&
	-2 \partial_{k}p^{kj}
	-2\partial_{k}(h_{ij}p^{ki})+p^{kl}\partial_{j}h_{kl} \,,
	\label{momentunconstpertur}
	\\
	\mathcal{H} &=&
	p^{ij}p^{ij}+\frac{\lambda}{1-d\lambda}p^{2} + \sqrt{g}\mathcal{V} \,.
\end{eqnarray}
We perform the transverse-longitudinal decomposition
\begin{equation}\label{decomposition}
	h_{ij} = h_{ij}^{TT}
	+\frac{1}{d-1}\left(\delta_{ij} - \frac{\partial_i\partial_j}{\Delta}\right)h^{T}
	+ \partial_{(i}h_{j)} \,,
\end{equation}
and similarly for $p^{ij}$. In $d=2$ dimensions the ${TT}$ mode must be absent from this decomposition. Thus, the action (\ref{actionpiijproj}) becomes
\begin{equation}
	\begin{split}
	S=&
	\int dt d^{d}x 
	\left[
	p_{TT}^{ij}\dot{h}_{ij}^{TT}
	+\frac{1}{(d-1)}p^{T}\dot{h}^{T}
	-\frac{1}{2}p^{i} \mathfrak{D}^{ij}_1 \dot{h}_{j}
	+\frac{1}{4} (\dot{n}^{k}-\Gamma^{k}) \mathfrak{D}^{-1}_{kl} (\dot{n}^{l}-\Gamma^{l}) 
	\right. \\ & 
	- p^{ij}_{TT}p^{ij}_{TT} 
	- \frac{1-\lambda}{(d-1)(1-d\lambda)}(p^{T})^{2}
	+ \frac{1}{2}p^{i}\left( \delta_{ij} \Delta +\left(1+\frac{2\lambda}{1-d\lambda}\right) \partial_i\partial_{j}\right) p^{j}
	\\ & \left.
	- \frac{2\lambda}{1-d\lambda}p^{T}\partial_ip^{i} 
	+ n^i \mathfrak{D}^{ij}_1 p^j
	- \sqrt{g}\mathcal{V}
	+\dot{C}^{k} \left( \dot{\bar{C}}_{k} 
	+\bar{C}_i \frac{\delta\Gamma^{i}}{\delta n^{k}} \right) -\bar{C}_i\{\Gamma^{i},\mathcal{H}_k\} C^{k}
	\right] \,.
	\end{split}
\label{actionprepijpertur}
\end{equation}
Note that the $(p^T)^2$ term disappears in the relativistic limit $\lambda = 1$, hence we assume that $\lambda$ does not take this value. By integrating $p^{ij}_{TT}$ and $p^{T}$, the action takes the form\footnote{Note that, since we are assuming that $\Gamma^k$ does not depend on $p^{ij}$, the last term in (\ref{actionprepijpertur}) does not depend on this momentum.} 
\begin{equation}
\begin{split}
	S=&
	\int\,dt\,d^{d}x\left\lbrace\frac{1}{4}\dot{h}_{ij}^{TT}\dot{h}_{ij}^{TT}+\frac{1-d\lambda}{4(1-\lambda)(d-1)}(\dot{h}^{T})^{2}
	+\frac{1}{4} (\dot{n}^{k}-\Gamma^{k}) \mathfrak{D}^{-1}_{kl} (\dot{n}^{l}-\Gamma^{l}) 
	\right. \\
	&	
	+\frac{\lambda}{1-\lambda}p^{i}\partial_{i}\dot{h}^{T}
	+ p^i \mathfrak{D}^{ij}_1 \left( -\frac{1}{2} \dot{h}_j + n^j \right)
	+\frac{1}{2} p^{i} \mathfrak{D}_2^{ij} p^{j} 
	- \sqrt{g}\mathcal{V}
	\\	&\left.
	+\dot{C}^{k} \left( \dot{\bar{C}}_{k} 
	+\bar{C}_i \frac{\delta\Gamma^{i}}{\delta n^{k}} \right) 
	-\bar{C}_l\{\Gamma^{l}\,,\mathcal{H}_k\} C^{k} \right\rbrace \,.
\end{split}
\label{actionprepilong}
\end{equation}
The last integration is on $p^{i}$. The square involving this variable can be completed,
\begin{equation}
  \frac{1}{2} \mathfrak{D}_2^{ij} 
  \left(p^{i} + \mathfrak{D}^{-1}_{2ik} B^{k}\right) 
 \left( p^{j} + \mathfrak{D}^{-1}_{2jl} B^{l} \right)
- \frac{1}{2} B^i \mathfrak{D}_{2ij}^{-1} B^j,
\end{equation}
where
\begin{align}
B^{k} =&
	\frac{\lambda}{1-\lambda}\partial_{k}\dot{h}^{T}
	- \mathfrak{D}^{kl}_1 \left(\frac{ \dot{h}_{l}}{2} - n^{l}\right) \,,
\label{B}
\\
		-\frac{1}{2}{B}^{k} \mathfrak{D}^{-1}_{2kl} {B}^{l} =&
		-\frac{1}{8}\dot{h}^{l} \mathfrak{D}^{kl}_3 \dot{h}_{k}
		+\frac{1}{2}\dot{h}_{l} \mathfrak{D}^{kl}_3 n^{k} 
		-\frac{1}{2}{n}^{l} \mathfrak{D}^{kl}_3 {n}^{k}
		\nonumber \\	&
		+\frac{\lambda^{2}}{4(1-\lambda)} (\dot{h}^{T})^2
		+\frac{1}{2}\lambda\dot{h}_k\partial_{k}\dot{h}^{T}
		-\lambda n^k\partial_{k}\dot{h}^{T} \,.
\end{align}
After the Gaussian integration, the action (\ref{actionprepilong}) becomes
\begin{equation}
	\begin{split}
	S =& 
	\int dt d^{d}x \left\lbrace \frac{1}{4}\dot{h}_{ij}^{TT}\dot{h}_{ij}^{TT}
	+\frac{(1-d\lambda)}{4(1-\lambda)(d-1)}(\dot{h}^{T})^{2}
	+\frac{1}{4} (\dot{n}^{k}-\Gamma^{k}) \mathfrak{D}^{-1}_{kl} (\dot{n}^{l}
	- \Gamma^{l})
	\right. \\
	&\left.
	- \sqrt{g}\mathcal{V}
	-\frac{1}{2} {B}^{k} \mathfrak{D}^{-1}_{2kl} {B}^{l}
	+ \dot{C}^{k} \left( \dot{\bar{C}}_{k} 
	+\bar{C}_i \frac{\delta\Gamma^{i}}{\delta n^{k}} \right) 
	- \bar{C}_i\{\Gamma^{i} \,, \mathcal{H}_k\} C^{k}
	\right\rbrace \,.
	\end{split}
\label{finalquantumddim}
\end{equation}

So far, the potential $\mathcal{V}$ and the factor $\Gamma^i$ of the gauge-fixing condition have been left unspecified, hence all the above formulas for projectable Ho\v{r}ava theory are valid in any spatial dimension $d$, except for the fact that in the $d=2$ case the $h_{ij}^{TT}$ mode must be dropped from all expressions. Now, to continue on obtaining the quantum Lagrangian, we specialize to the $d=2$ case, specifying the potential and the gauge-fixing condition completely. The potential of the $d=2$ projectable Ho\v{r}ava theory, up to second order in perturbations, becomes
\begin{equation}
	\sqrt{g} \mathcal{V} = \mu\sqrt{g}R^{2} = \mu  ( \Delta h^{T} )^2 \,.
\end{equation} 
The operator $\mathfrak{D}^{ij}$ is defined in (\ref{D}). For the factor $\Gamma^i$ we take the form introduced in Ref.~\cite{Barvinsky:2015kil}, which was obtained by considering the anisotropic scaling of the variables of the Ho\v{r}ava theory,
\begin{equation}
	\Gamma^{k} =
	  2 c_{1} \Delta\partial_lh_{kl}
	+ 2 c_{2} \Delta\partial_{k}h
	+ c_{3} \partial_{k}\partial_{i}\partial_{j} h_{ij} \,,
\label{gammaproj}
\end{equation}
where $c_{1},c_{2},c_{3}$ are constants. In the transverse--longitudinal decomposition it takes the form
\begin{equation}
	\Gamma^{k} =
	c_{1}\Delta^2 h_{k}
	+\gamma \Delta \partial_{k}\partial_{l} h_l
	+ 2 c_{2} \Delta \partial_{k}h^{T} \,,
\end{equation}
where  $\gamma=c_{1} + 2 c_{2} + c_{3}$. Now we may write explicitly several elements of the action (\ref{finalquantumddim}) for the $d=2$ case. We have the terms that involve the time derivative of the shift vector,
\begin{equation}
	\begin{split}
	&\frac{1}{4}(\dot{n}^k-\Gamma^k) \mathfrak{D}^{-1}_{kl} (\dot{n}^l-\Gamma^l) =
	\frac{1}{4}\dot{n}^l \mathfrak{D}^{-1}_{kl} \dot{n}^k 
	-\frac{1}{2}\dot{n}^{k}\Big(
	c_{1} \left(\delta_{kl}\Delta 
	- \rho \kappa \partial_k\partial_l \right)
	+ \rho \gamma \partial_k\partial_l\Big)
	h_{l} \\
	&
	- \rho c_{2} \dot{n}^{k}\partial_kh^{T}
	+\frac{1}{4}h_{k}\left[ c_{1}^{2} \delta_{lk} \Delta^3
	+\left( 2 c_{1}\gamma + \gamma^{2}
	- \rho\kappa \left( \gamma + c_1 \right)^{2}\right)
	\Delta^2 \partial_k\partial_l\right]h_{l}
	\\ &
	+ \rho c_{2} ( \gamma + c_1 )  h_{k}
	\Delta^2 \partial_k h^{T}
	- \rho c_{2}^{2} h^{T} \Delta^2 h^{T} \,,
	\end{split}
\end{equation}
where $\rho = (1+\kappa)^{-1}$. In the ghost sector we have the bracket
\begin{equation}
	\int d^{2}x \bar{C}_i  \{\Gamma^{i} \,, \mathcal{H}_k \} C^{k} =
	- 2 \int\,d^{2}x
	C^{k}( c_{1} \delta_{kl} \Delta^2 + \gamma\Delta \partial_k\partial_{l})\bar{C}_l \,.
\end{equation}
The action in $2+1$ dimensions takes the form 
\begin{equation}\label{actiongeneral1}
	\begin{split}
S =&
\int\,dt\,d^{2}x\left\lbrace
\frac{1-\lambda}{4} (\dot{h}^{T})^{2}
-\left( \mu + \rho c_{2}^{2} \right) h^{T}\Delta^2 h^{T}
+\frac{1}{4}\dot{n}^l \mathfrak{D}^{-1}_{kl} \dot{n}^k
+\frac{\lambda}{2}\dot{h}_{k}\partial_k\dot{h}^{T}
\right.\\
&
- \frac{1}{8}\dot{h}_{l} \mathfrak{D}^{kl}_3 \dot{h}_{k}
+ \rho c_{2} ( \gamma + c_1 ) h_{k}\Delta^2 \partial_kh^{T}
-\frac{1}{2}n^{l} \mathfrak{D}^{kl}_3 n^{k}
+ \left( \lambda - \rho c_2 \right)    
   \dot{n}^{k}\partial_kh^{T} 
\\ &
- \frac{1}{2}\dot{n}^{k}\left[ 
    \left( 1 + c_{1} \right) \delta_{lk} \Delta 
    + \left(  1 - 2\lambda + \rho ( \gamma - \kappa c_{1} ) \right) \partial_k\partial_l\right]h_{l}
\\  &
+\frac{1}{4}h_{k}\left[ c_{1}^{2} \delta_{kl}\Delta^3
+\left( 2 c_{1}\gamma + \gamma^{2}
- \rho \kappa \left( \gamma + c_1 \right)^{2}\right)
\Delta^2 \partial_k\partial_l\right]h_{l}
\\ & \left.
+ \dot{C}^{k}\dot{\bar{C}}_k
+ 2 C^{k}( c_{1}\delta_{kl}\Delta^2 + \gamma\Delta \partial_k\partial_{l})\bar{C}_l 
\right\rbrace.
\end{split}
\end{equation}
We notice the presence of odd derivatives in time or space in (\ref{actiongeneral1}), which are also the terms that mix $n^i$ and the components of $h_{ij}$\footnote{ Actually, the odd derivatives are only in time. We see this explicitly in the $3+1$ nonprojectable case, where we present the analogous formulas, Eqs.~(\ref{BB}) and (\ref{gaugeS}), written with the decomposition on the vectors fields.}. We see that these odd terms cancel if we set\footnote{This identification is equivalent to the one used in Ref.~\cite{Barvinsky:2015kil} for the Lagrangian theory.} 
\begin{equation}
 c_{1}= - 1 \,,
 \qquad 
 c_{2}= \lambda(1+\kappa) \,,
 \qquad 
 c_{3}=-2\kappa \,.	
\label{settingconstants}
\end{equation}
By adjusting these constants, the final quantum path integral of the projectable $2+1$ Ho\v{r}ava theory, written in Lagrangian variables and at second order in perturbations, is
\begin{equation}
\begin{split}
 Z =& 
  \int \mathcal{D} h^T \mathcal{D} h_i^T \mathcal{D} h^L
  \mathcal{D} n^{iT} \mathcal{D} n^L \mathcal{D} \bar{C}_i \mathcal{D} C^i
  \\ &
  \exp \left[ i 	
	\int dt d^{2}x \left[h^{T}\Big(-\frac{1}{4}(1-\lambda)\partial_t^{2}
	-\left(\mu+(1+\kappa)\lambda^{2}\right)\Delta^2 \Big)h^{T}
	\right. \right. \\
	& +C^{k}\left[-\partial_t^{2}
	-2\delta_{kl} \Delta^2 +2 \left(2(\lambda-1)(1+\kappa)+1\right)
	\Delta\partial_k\partial_{l}\right]\bar{C}_l  \\
	& -\frac{1}{4}n^{kT} \Big(\Delta^{-1}\partial_t^{2}+2\Delta\Big)n^{kT}
	+n^L\left(\frac{1}{4(1+\kappa)}\partial_t^{2}+(1-\lambda)\Delta^2 \right)n^L 
	\\
	& +h^{T}\Big(\frac{1}{2}\lambda\partial_t^{2}-2\lambda(\lambda-1)(1+\kappa)\Delta^2 \Big) h^L
	+\frac{1}{8}{h}_k^{T}\Big(\Delta\partial_t^{2}
	+2\Delta^3\Big)h^{k}{}^{T} \\
	& \left.
	-h^L\Big(\frac{1}{4}(1-\lambda)\partial_t^{2}+(\lambda-1)^{2}(1+\kappa) \Delta^2 \Big) h^L \right] \,,
	\end{split}
\label{finalprojectable}
\end{equation}
where we have also decomposed the vectors, 
\begin{equation}
	\begin{split}
		& n^{i}=n^{i}{}^{T}+\partial^{i}n^L \,, 
		\qquad \partial_in^{i}{}^{T}=0 \,,
		\\ &
		h_{i}=
		h_{i}{}^{T}+\partial_{i}\Delta^{-1}h^L \,, 
		\qquad \partial_i h_{i}{}^{T}=0 \,.
	\end{split}
	\label{vectordecomposition}
\end{equation}
The quantum Lagrangian of Eq.~(\ref{finalprojectable}) coincides with the one presented in Ref.~\cite{Barvinsky:2015kil}. Those authors used a Faddeev-Popov procedure for fixing the gauge, hence they get the usual parameter $\sigma$ associated to the averaging on the gauge-fixing condition. To match exactly both Lagrangians, we must set $\sigma = 1/4$. At the end, the nonlocality only affects the time-derivative of the shift vector (and all propagators are regular \cite{Barvinsky:2015kil}).

Finally, we make a comment on the cubic order in perturbations in the ghost sector. We take the ghost sector of the action given in (\ref{prepiij}). Its expansion up to cubic order, imposing the gauge (\ref{gammaproj}), is 
\begin{equation}
\begin{split}
	S_{gh}=&
	\int dt d^{2}x \left\{
	  \dot{C}^{k}\dot{\bar{C}}_{k}
	+ \dot{\bar{C}}_k\partial_iC^{k}n^{i}
	- C^{k} \partial_i \dot{\bar{C}}_k n^{i} 	
	- \dot{\bar{C}}_kC^{i}\partial_in^{k}
	+ C^{i} \left[ -\delta_{ij} \Delta^2
	\right.\right. \\ & \left.
	+ ( 2 \lambda -1 + 2\kappa(\lambda-1) ) \Delta \partial_{i} \partial_{j} \right] \bar{C}_j 
	- \Delta\partial_k\bar{C}_l ( \partial_{k} C^j h_{lj}
	+ \partial_{l}C^j h_{kj}
	+ C^j\partial_{j} h^{kl} )
	\\ & \left.
	+\lambda(1+\kappa)\Delta\partial_{i}\bar{C}_i
	(2\partial_{l}C^{k} h_{lk}+\partial_khC^{k})
	-\kappa\partial_{k}\partial_{l}\partial_{j} \bar{C}_j (2\partial_kC^{i}h_{li}+\partial_ih_{kl}C^{i}) \right\} .
	\end{split}
\end{equation}
This is equal to the cubic order in the ghost sector of Ref.~\cite{Barvinsky:2015kil}, except for an additional term we find, which is $- C^{k} \partial_i \dot{\bar{C}}_k n^{i}$.


\section{$3+1$ Nonprojectable theory}
\subsection{Classical theory}
In the nonprojectable theory the lapse function $N$ is allowed to depend on time and space, hence it represents a complete functional degree of freedom. In this case a large class of terms that depend on the vector $a_i = \partial_i \ln N$ arise in the Lagrangian \cite{Blas:2009qj}. We focus the nonprojectable theory in $3+1$ dimensions. The Lagrangian has the general form shown in (\ref{lagrangian}), but eliminating the restriction of projectability on $N$. The criterium of power-countig renormalizability requires us to include a term of order $z=3$ in $3+1$ dimensions. The total Lagrangian, containing the $z=1,2,3$ orders has many terms. In this analysis we take for the potential only the $z=3$ terms that contribute to the propagators, which are the dominant terms in the propagators in the ultraviolet regime. They are \cite{Colombo:2014lta}
\begin{equation}
	\mathcal{V}=
	-\alpha_{3}\nabla^{2}R\nabla_{i}a^{i}-\alpha_{4}\nabla^{2}a_{i}\nabla^{2}a^{i}-\beta_{3}\nabla_{i}R_{jk}\nabla^{i}R^{jk}-\beta_{4}\nabla_{i}R\nabla^{i}R \,,
\end{equation}
where $\alpha_3,\alpha_4,\beta_3,\beta_4$ are coupling constants.

In the nonprojectable theory the lapse function $N$ and its conjugate momentum $P_N$ are part of the canonical variables. There is no time derivative of $N$ in the Lagrangian, hence 
\begin{equation}
 \theta_1 \equiv P_N = 0
\end{equation}
is a constraint of the theory. The classical Hamiltonian, obtained by a Legendre transformation, is
\begin{equation}
	H_{0}=\int d^{3}x\sqrt{g}N\left(\frac{\pi^{ij}\pi_{ij}}{g}+\frac{\lambda}{1-3\lambda}\frac{\pi^{2}}{g}+\mathcal{V}\right) \,.
\label{H0}
\end{equation}
The rest of constraints are the momentum constraint, $\mathcal{H}_{i}=-2\nabla^k\pi_{ki}$, and the constraint
\begin{equation}
\begin{split}
 \theta_2 \equiv & \frac{N}{\sqrt{g}}\left(\pi^{ij}\pi_{ij}
 +\frac{\lambda}{1-3\lambda}\pi^{2}\right)
 +\sqrt{g} N \mathcal{V}	
 -\alpha_{3}\sqrt{g} \nabla^{2}(N\nabla^{2}R)
\\ &
 +2\alpha_{4} \sqrt{g} \left(\nabla^{i}\nabla^{2}(N\nabla^{2}a_{i})\right) = 0\,.
\end{split}
\label{hamiltonianconst}
\end{equation}
In the definition of the phase space, the main qualitative difference between the projectable and nonprojectable cases is the activation of the lapse function as a degree of freedom and the arising of the constraint $\theta_2$ (\ref{hamiltonianconst}) in the side of the nonprojectable theory. The last two terms of the constraint $\theta_2$ are total derivatives of sixth order, hence the integral of $\theta_2$ is equal to the primary Hamiltonian (\ref{H0}),
\begin{equation}\label{eq8}
 H_0 = \int d^3x \,\theta_2 \,.
\end{equation}
Actually, when the $z=1$ terms are included in the potential, there is a boundary contribution remaining from the integral of $\theta_2$. Moreover, a term proportional to the so called Arnowitt-Deser-Misner energy is required for the differentiability of one of the $z=1$ terms. Therefore, the general statement is that the primary Hamiltonian of the $3+1$ nonprojectable Ho\v{r}ava theory can be written as the integral of $\theta_2$ plus boundary terms. Since in this analysis we focus on the $z=3$ terms, we can discard these boundary terms.


\subsection{BFV quantization}
Since the nonprojectable theory has second-class constraints, the definitions of the BFV quantization must be adapted, according to Ref.~\cite{Fradkin:1977xi}. The involution is defined in terms of Dirac brackets,
\begin{equation}
	\left\lbrace G_{a},G_{b}\right\rbrace_{D}=U^{c}_{ab}G_{c},\;\;\;\left\lbrace\mathcal{H}_{0},G_{a}\right\rbrace_{D}=V^{b}_{a}G_{b} \,,
\end{equation}
where Dirac brackets are defined by
\begin{equation}
	\left\lbrace F,R\right\rbrace_{D}=\left\lbrace F,R\right\rbrace-\left\lbrace F,\theta_{A}\right\rbrace\mathbb{M}^{-1}_{AB}\left\lbrace\theta_{B},R\right\rbrace,\;\;\;\mathbb{M}_{AB}=\left\lbrace\theta_{A},\theta_{B}\right\rbrace.
\end{equation}

The implementation of the BFV quantization of the $3+1$ case is parallel to the $2+1$ case shown in Ref.~\cite{Bellorin:2021tkk}. Here we present the summary. The matrix of Poisson brackets of the second-class constraints has a triangular form,
\begin{equation}\label{matrixM}
	\mathbb{M}=\begin{pmatrix}
		0 & \lbrace\theta_{1},\theta_{2}\rbrace \\
		-\lbrace\theta_{1},\theta_{2}\rbrace & \lbrace\theta_{2},\theta_{2}\rbrace
	\end{pmatrix} \,.
\end{equation}
Since the primary Hamiltonian $\mathcal{H}_0$ is equivalent to the second-class constraint $\theta_2$, its Dirac bracket is zero with any quantity, hence $V_a^b = 0$. The Dirac bracket of the momentum constraint $\mathcal{H}_i$ with itself is equivalent to its Poisson bracket,
\begin{equation}
	\left\lbrace \mathcal{H}_{i}(x),\mathcal{H}_{j}(y)\right\rbrace_{D}=
	\left\lbrace \mathcal{H}_{i}(x),\mathcal{H}_{j}(y)\right\rbrace\label{HH} \,,
\end{equation}
This leads to the algebra of spatial diffeomorphisms, as in the projectable case, hence the coefficients $U_{ij}^k$ are the same, and $U_{ab}^c = 0$ for $a,b,c > i$.

We perform the BFV extension of the phase space in a similar way to the projectable case. The Lagrange multipliers form a new canonical pair $(N^{i},\pi_{i})$. The ghosts are the canonical pairs $(\eta^{a},\mathcal{P}_{a})$. Thus, the full phase space is given by the pairs $(g_{ij},\pi^{ij})$, $(N,P_{N})$, $(N_{i},\pi^{i})$ and $(\eta_{a},\mathcal{P}^{a})$. The BFV path integral of the nonprojectable Ho\v rava theory is given by
\begin{equation}
	Z = 	\int\mathcal{D}V\delta(\theta_{1})\delta(\theta_{2})e^{iS},
\end{equation}
where the measure and the action are given by
\begin{align}
	\mathcal{D}V&=
	\mathcal{D}g_{ij}\mathcal{D}\pi^{ij}\mathcal{D}N\mathcal{D}P_{N}\mathcal{D}N^{k}\mathcal{D}\pi_{k}\mathcal{D}\eta^{a}\mathcal{D}\mathcal{P}_{a}\times\sqrt{\det\mathbb{M}} \,,
	\\
	S&=\int dtd^{3}x\left(\pi^{ij}\dot{g}_{ij}+P_{N}\dot{N}+\pi_{k}\dot{N}^{k}+\mathcal{P}_{a}\dot{\eta}^{a}-\mathcal{H}_{\Psi}\right).\label{principalaction}
\end{align}
Unlike the projectable case, in this case the second-class constraints must be imposed explicitly. By comparing the quantum degrees of freedom with the projectable case, here we see that the canonical pair $(N,P_{N})$ has been added to the phase space, but at the same time the imposition of the two second-class constraints $\theta_1,\theta_2$ compensates the pair $(N,P_{N})$. Hence the balance is the same of the projectable case: there are $d(d-1)$ physical degrees of freedom in the quantum phase space. In the $3+1$ foliation, they are the two tensorial modes and the extra scalar mode.  

The BRST charge $\Omega$ and $\mathcal{H}_1$ take the same definition of Eqs.~(\ref{omega}) and (\ref{H1}). The quantum gauge-fixed Hamiltonian is
\begin{equation}
	\mathcal{H}_{\Psi}=\mathcal{H}_{1}+\left\lbrace\Psi,\Omega\right\rbrace_{D}\,.
\end{equation}
The nonprojectable Ho\v rava theory is a theory of rank one, then
\begin{align}
	\Omega &=
	G_{a}\eta^{a}-\frac{1}{2}U^{c}_{ab}\eta^{a}\eta^{b}\mathcal{P}_{c}=\mathcal{H}_{k}\eta^{k}_{1}+\pi_{k}\eta^{k}_{2}-\frac{1}{2}U^{k}_{ij}\eta^{i}_{1}\eta^{j}_{1}\mathcal{P}^{1}_{k} \,,
	\\
	\mathcal{H}_{1} &=
	\mathcal{H}_{0} \,.
\end{align}
In the case of the second-class constraints the consistency conditions for the BFV quantization are
\begin{equation}
 \{ \Omega \,, \Omega \}_D = 0 \,,
 \qquad
 \{ \mathcal{H}_1 \,, \Omega \}_D = 0 \,.
\end{equation}
The first condition holds following the same steps of the projectable case, but operating with Dirac brackets in this case. The second condition holds because $\mathcal{H}_1 = \mathcal{H}_0$, and this is equivalent to a second-class constraint, hence its Dirac bracket is always zero. The gauge-fixed quantum Hamiltonian takes the form
\begin{equation}\label{HBRST}
	\mathcal{H}_{\Psi}=\mathcal{H}_{0}+\left\lbrace\Psi,\mathcal{H}_{k}\eta_{1}^{k}+\pi_{k}\eta^{k}_{2}-\frac{1}{2}U^{k}_{ij}\eta^{i}_{1}\eta^{j}_{1}\mathcal{P}_{k}^{1}\right\rbrace_{D}.
\end{equation}

As we did in the projectable case, we can adopt the form of the gauge-fixing condition used in the general BFV formalism, originally introduced for relativistic theories. Thus, the gauge-fixing condition $\Phi^i = 0$ and the associated fermionic function $\Psi$ take the forms given in (\ref{gaugefixing}) and (\ref{Psi}), respectively. The Hamiltonian takes the form 
\begin{equation}
\begin{split}
	{\mathcal{H}}_{\Psi} =&
	\mathcal{H}_{0} 
	+\mathcal{H}_kN^{k}
	+ \mathcal{P}_k^{1} \eta_2^{k}
    - \mathcal{P}^{1}_{i}\left(N^{j}\partial_{j}\eta_{1}^{i} + N^{i}\partial_{j}\eta_{1}^{j}\right)
	+\pi_k\chi^{k}
	\\ &
	+ \mathcal{P}_i^{2}\{\chi^{i} \,, \mathcal{H}_k\} \eta_1^{k}
	+ \mathcal{P}_i^{2} \dfrac{ \delta\chi^{i}}{\delta N^{l}} \eta_2^{l} \,.
\end{split}
\end{equation}

Due to the form (\ref{matrixM}), the measure of the second-class constraints simplifies to $\sqrt{\det\mathbb{M}} = \det \{ \theta_1 \,, \theta_2 \}$. Thus, this measure can be incorporated to the Lagrangian by means of the ghosts fields $\bar{\varepsilon},\varepsilon$,
\begin{equation}
	\det\left\lbrace\theta_{1},\theta_{2}\right\rbrace=\int\mathcal{D}\bar{\varepsilon}\mathcal{D}\varepsilon\exp\left(i\int dtd^{3}x\;\bar{\varepsilon}\left\lbrace\theta_{1},\theta_{2}\right\rbrace\varepsilon\right).
\end{equation}
Taking the definition of $\theta_2$ given in (\ref{hamiltonianconst}), the bracket $\{ \theta_1 \,, \theta_2 \}$ results
\begin{equation}
	\begin{split}
		\left\lbrace\theta_{1}(x),\theta_{2}(y)\right\rbrace &=
		\frac{\theta_2}{N} \delta_{xy}
		+2 \sqrt{g}\left\lbrace
		  - \alpha_{3}\nabla^{i}\left(\nabla_{i}\delta_{xy} \nabla^{2}R
		  -	\delta_{xy} a_{i}\nabla^{2}R\right)
		\right.\\
		&\left.
		+ \alpha_{4}\left[\nabla_{i}\nabla^{2}(\delta_{xy}\nabla^{2}a^{i})
		- N\nabla^{2}a^{i}\nabla^{2}\nabla_{i}\left(\frac	{\delta_{xy}}{N}\right)
		\right.\right. 
	    \\ & \left.\left.
	    +\nabla_{i}\nabla^{2}\left(N\nabla^{2}\nabla^	{i}\left(\frac{\delta_{xy}}{N}\right)\right)
	    -\frac{\delta_{xy}}{N}\nabla_	{i}\nabla^{2}(N\nabla^{2}a^{i})\right]\right\rbrace \,,
	\end{split}
	\label{measure}
\end{equation}
where $\delta_{xy} \equiv \delta(x^i - y^i)$. Once we have obtained this bracket, we may integrate the variable $P_N$ without further consequences, since it vanishes due to the constraint $\theta_1 = 0$. The constraint $\theta_{2}$ can be incorporated to the Lagrangian by means of a Lagrange multiplier, which we denote by $\xi$. Thus, the BFV path integral of the nonprojectable Ho\v{r}ava theory in $3+1$ dimensions takes the form
\begin{equation}
	\begin{split}
		Z_{\Psi} =& 
		\int \mathcal{D} g_{ij} \mathcal{D}\pi^{ij} \mathcal{D}N \mathcal{D}N^{k} \mathcal{D}\pi_{k} \mathcal{D}\eta_1^i \mathcal{D}\mathcal{P}^1_i \mathcal{D}\eta_2^i \mathcal{D}\mathcal{P}^2_i
		\mathcal{D}\xi
		\mathcal{D}\bar{\varepsilon}
		\mathcal{D}\varepsilon
		\\ &
		\exp\left[ i \int dt d^{3}x 
		\Big(\pi^{ij}\dot{g_{ij}} +\pi_{k}\dot{N}^{k}
		+\mathcal{P}^1_i \dot{\eta}_1^i + \mathcal{P}^2_i \dot{\eta}_2^i
		- \mathcal{H}_{0} 
		- \mathcal{H}_kN^{k}
		- \mathcal{P}_k^{1} \eta_2^{k}
		\right. 
		\\ &  
		+ \mathcal{P}^{1}_{i}\left(N^{j}\partial_{j}\eta_{1}^{i}+N^{i}\partial_{j}\eta_{1}^{j}\right)
		- \pi_k\chi^{k}
		- \mathcal{P}_i^{2}\{\chi^{i} \,, \mathcal{H}_k\} \eta_1^{k}
		- \mathcal{P}_i^{2} \dfrac{ \delta\chi^{i}}{\delta N^{l}} \eta_2^{l}
		\\ & \left.
		+ \xi\theta_{2}
		+ \bar{\varepsilon} \left\lbrace\theta_{1} \,, \theta_{2}\right\rbrace \varepsilon
		\Big)\right]
		\,.
	\end{split}
\end{equation}

\subsection{Quantum Lagrangian}
By adapting the discussion done in section 2 about the structure of the gauge-fixing condition to the nonprojectable case, we set
\begin{equation}\label{gauge1}
	\chi^{i}= 
	\mathfrak{D}^{ij}\pi_{j}
	+\Gamma^{i}[g_{ij},N,N^i] \,.
\end{equation}
The Gaussian integration on $\pi_i$ leads to the path integral
\begin{equation}
	\begin{split}
		Z_{\Psi} =& 
		\int \mathcal{D}g_{ij} \mathcal{D}\pi^{ij} \mathcal{D}N \mathcal{D}N^{k} \mathcal{D}\eta_1^i \mathcal{D}\mathcal{P}^1_i \mathcal{D}\eta_2^i \mathcal{D}\mathcal{P}^2_i
		\mathcal{D}\xi
		\mathcal{D}\bar{\varepsilon}
		\mathcal{D}\varepsilon
		\\ &
		\exp\left[ i \int dt d^{3}x 
		\Big(\pi^{ij}\dot{g_{ij}}
		+\mathcal{P}^1_i \dot{\eta}_1^i + \mathcal{P}^2_i \dot{\eta}_2^i
		+ \frac{1}{4} (\dot{N}^{k}-\Gamma^{k}) \mathfrak{D}^{-1}_{kl} (\dot{N}^{l}-\Gamma^{l})
		\right.
		\\ &	
		- \mathcal{H}_{0} 
		- \mathcal{H}_kN^{k}
		- \mathcal{P}_k^{1} \eta_2^{k}
		+ \mathcal{P}^{1}_{i}\left(N^{j}\partial_{j}\eta_{1}^{i}+N^{i}\partial_{j}\eta_{1}^{j}\right)  
		\\ & \left.
		- \mathcal{P}_i^{2}\{\Gamma^{i} \,, \mathcal{H}_k\} \eta_1^{k}
		- \mathcal{P}_i^{2} \dfrac{ \delta\Gamma^{i}}{\delta N^{l}} \eta_2^{l}
		+\xi\theta_{2}+\bar{\varepsilon}\left\lbrace\theta_{1},\theta_{2}\right\rbrace\varepsilon\Big)\right]
		\,.
	\end{split}
\end{equation}

The next integration we perform is over the BFV ghosts that are canonical momenta. We perform the same change of notation (\ref{newnotation}). For the terms of the action that depend on $\mathcal{P}^{a}$ and $\bar{\mathcal{P}}_{a}$ it is possible to carry out the integration after completing the bilinear in these variables, as in (\ref{squaregrassmann}). The action of the ghost sector results
\begin{equation}
	\begin{split}
		S_{\text{ghost}}&=\int dtd^{3}x\left\lbrace\left(\dot{C}^{k}-N^{j}\partial_{j}C^{k}-N^{k}\partial_{j}C^{j}\right)\left(\dot{\bar{C}}_k+\bar{C}_i\dfrac{ \delta\Gamma^{i}}{\delta N^{k}}\right)
		- \bar{C}_{i} \left\lbrace\Gamma^{i} \,, \mathcal{H}_{k} \right\rbrace_{D} C^{k} \right\rbrace.
	\end{split}
\end{equation}

Now we adopt the perturbative variables defined in (\ref{perturbations}), adding $N-1 = n$. For the $d=3$ nonprojectable theory we take \cite{Barvinsky:2015kil}
\begin{equation}
	\mathfrak{D}^{ij} = \delta_{ij} \Delta^2 
	+ \kappa \Delta \partial_i \partial_j \,,
	\quad
	\mathfrak{D}^{-1}_{ij}=\frac{\delta_{ij}}{\Delta^{2}}-\frac{\kappa}{1+\kappa}\frac{\partial_{i}\partial_{j}}{\Delta^{3}} \,.
\end{equation}
The momentum constraint $\mathcal{H}_{j}$ is given in (\ref{momentunconstpertur}), and the Hamiltonian density $\mathcal{H}_{0}$ takes the form
\begin{equation}
	\begin{split}
	\mathcal{H}_{0} =&
	p^{ij}p^{ij}+\frac{\lambda}{1-3\lambda}p^{2}-\alpha_{3}n\Delta^{2}\left(\partial_{i}\partial_{j}h_{ij}-\Delta h\right)
	+\alpha_{4}n\Delta^{3}n
    \\ & +\left(\frac{\beta_{3}}{2}+\beta_{4}\right)h_{kj}\partial_{j}\partial_{k}\partial_{n}\partial_{l}\Delta h_{nl} +\left(\frac{\beta_{3}}{4}+\beta_{4}\right)\left(h\Delta^{3}h-2h\partial_{j}\partial_{k}\Delta^{2}h_{jk}\right)
   \\ &
	+ \frac{\beta_{3}}{4} \left( h_{kj}\Delta^{3}h_{kj} - 2h_{kj}\partial_{l}\partial_{k}\Delta^{2}h_{lj}\right).
\end{split}
\end{equation}
Therefore, the path integral becomes
\begin{equation}
	\begin{split}
		Z = &
		\int \mathcal{D} h_{ij} \mathcal{D} p^{ij} \mathcal{D}n \mathcal{D}n^{k} \mathcal{D} \bar{C}_i \mathcal{D} C_i 
		\mathcal{D}\bar{\varepsilon} \mathcal{D}\varepsilon
		\mathcal{D}\xi
		\\ &
		\exp \left[ i
		\int dtd^{3}x \left( p^{ij}\dot{h}_{ij}-\mathcal{H}_{0}
		-n^{k}\mathcal{H}_{k}
		+\frac{1}{4} (\dot{n}^{k}-\Gamma^{k}) \mathfrak{D}^{-1}_{kl} (\dot{n}^{l}-\Gamma^{l})
		\right.\right. \\ & \left. \left.
		+\xi\theta_{2}
		+\bar{\varepsilon}\left\lbrace\theta_{1},
		\theta_{2}\right\rbrace\varepsilon
		+\dot{C}^{k}\left(\dot{\bar{C}}_k+\bar{C}_i\frac{\delta\Gamma^{i}}{\delta n^{k}}\right)
		- \bar{C}_{i} \left\lbrace\Gamma^{i},\mathcal{H}_{k} \right\rbrace_{D} C^{k} \right) \right] \,.
	\end{split}
\end{equation}
We make the decomposition (\ref{decomposition}) on the fields. The second class constraint $\theta_{2}$ and the measure of the second-class constraints, given by the bracket (\ref{measure}), contribute to the perturbative action with the following terms, respectively,
\begin{equation}
	\xi\theta_{2}=
	\xi \left( \alpha_{3} \Delta^{3}h^{T} + 2\alpha_{4} \Delta^{3}n \right) \,,\quad\bar{\varepsilon}\left\lbrace\theta_{1},\theta_{2}\right\rbrace\varepsilon=-2\alpha_{4}\bar{\varepsilon}\Delta^{3}\varepsilon \,,
\end{equation}
where the Lagrange multiplier $\xi$ is regarded as a perturbative variable. After these steps, the Gaussian integration on $p^{ij}_{TT}$ and $p^{T}$ can be done by completing squares (assuming again $\lambda \neq 1$). This yields the action
\begin{equation}
\begin{split}
	S &=
	\int dtd^{3}x\left\lbrace \frac{1}{4}\dot{h}_{ij}^{TT}\dot{h}_{ij}^{TT}+\frac{(1-3\lambda)}{8(1-\lambda)}(\dot{h}^{T})^2
	+ \frac{1}{4} \left( \dot{n}^{i}-\Gamma^{i}\right) \mathfrak{D}^{-1}_{ij} \left(\dot{n}^{j}-\Gamma^{j}\right)\right.\\
	&
	\left.
	+\frac{\lambda}{1-\lambda}p^{i}\partial_{i}\dot{h}^{T}
	+ p^{l} \mathfrak{D}_1^{kl}	\left( -\frac{1}{2} \dot{h}_{k} + n^k \right)
	+\frac{1}{2} p^{i} \mathfrak{D}_2^{ij} p^{j}
	-\alpha_{3}n\Delta^{3}h^{T} 
	-\alpha_{4}n\Delta^{3}n
	\right.\\ & \left.
	-\frac{\beta_{3}}{4}h^{TT}_{ij}\Delta^{3}h^{TT}_{ij}-\left(\frac{3\beta_{3}}{8}+\beta_{4}\right)h^{T}\Delta^{3}h^{T}
	+ \alpha_{3}\xi\Delta^{3}h^{T}+2\alpha_{4}\xi\Delta^{3}n
	\right. \\ & \left.
	+ \dot{C}^{k}\left(\dot{\bar{C}}_k+\bar{C}_i\frac{\delta\Gamma^{i}}{\delta n^{k}}\right)-2\bar{C}_{l}\frac{\delta\Gamma^{l}}{\delta h_{ij}} \partial_{j}C^i
	-2\alpha_{4}\bar{\varepsilon}\Delta^{3}\varepsilon
	\right\rbrace \,,
\end{split}
\end{equation}
where the operators $\mathfrak{D}_1^{ij}$ and $\mathfrak{D}_2^{ij}$ are the same of the $d=2$ case defined in (\ref{D1}) and (\ref{hatD}). The last integration is on $p^{i}$. We integrate in a similar way to how it was done in the projectable case, obtaining
\begin{equation}
	\begin{split}
		S&=
		\int dtd^{3}x\left\lbrace 
		\frac{1}{4}\dot{h}_{ij}^{TT}\dot{h}_{ij}^{TT}
		+\frac{(1-3\lambda)}{8(1-\lambda)}(\dot{h}^{T})^2	
		+\frac{1}{4} \left( \dot{n}^{i} - \Gamma^{i}\right) \mathfrak{D}^{-1}_{ij} \left(\dot{n}^{j}-\Gamma^{j}\right)
		\right.
		\\ 	&\left.
		-\frac{\beta_{3}}{4}h^{TT}_{ij}\Delta^{3}h^{TT}_{ij}
		-\left(\frac{3\beta_{3}}{8}+\beta_{4}\right)h^{T}\Delta^{3}h^{T}-\alpha_{3}n\Delta^{3}h^{T}-\alpha_{4}n\Delta^{3}n 
		+ \alpha_{3}\xi\Delta^{3}h^{T} \right.
		\\	&\left.
		+ 2\alpha_{4}\xi\Delta^{3}n
		- \frac{1}{2} B^{j} \mathfrak{D}_{2jl}^{-1} B^{l}
		+ \dot{C}^{k}\left(\dot{\bar{C}}_k+\bar{C}_i\frac{\delta\Gamma^{i}}{\delta n^{k}}\right)
		-2\bar{C}_{l}\frac{\delta\Gamma^{l}}{\delta h_{ij}} \partial_{j}C^i
		-2\alpha_{4}\bar{\varepsilon}\Delta^{3}\varepsilon
		\right\rbrace \,,
	\end{split}\label{actionnonprojectable}
\end{equation}
where $B^i$ is defined in Eq.~(\ref{B}). Finally, we make the decomposition on the vector variables shown in (\ref{vectordecomposition}). In particular,
\begin{equation}
	\begin{split}
	&
	- \frac{1}{2}B^{i} \mathfrak{D}_{2ij}^{-1}B^{j} =
	-\frac{\lambda^{2}}{4(1-\lambda)}h^{T}\partial^{2}_{t}h^{T}
    +\frac{\lambda}{2}h^{T}\partial_{t}^{2}h^{L}
	-\frac{1}{2}\dot{n}^{iT}\Delta h_i^{T}-\lambda\dot{n}^{L}\Delta h^{T}
	\\ &
	+(1-\lambda)\dot{n}^{L}\Delta h^{L}
	+\frac{1}{8}h_i^{T}\partial^{2}_{t}\Delta h_i^{T}-\frac{1}{2}n^{iT}\Delta n^{iT}
	-\frac{(1-\lambda)}{4}h^{L}\partial^{2}_{t}h^{L}
	+(1-\lambda)n^{L}\Delta^{2}n^{L} \,.
	\end{split}
\label{BB}
\end{equation}

Now we define the factor $\Gamma^i$ of the gauge-fixing condition, adopting the analysis of Ref.~\cite{Barvinsky:2015kil}. Those authors found that the appropriate gauge fixing condition in $d=3$, preserving the anisotropy of the Ho\v{r}ava theory, is given by
\begin{equation}
\Gamma^{i}=
  2 c_{1} \Delta^{2}\partial_{j}h_{ij}
+ 2 c_{2} \Delta^{2}\partial_{i}h
+ c_{3}\Delta\partial_{i}\partial_{j}\partial_{k}h_{jk} \,.
\end{equation}
The notation on the constants $c_{1,2,3}$ has been put intentionally equal to the projectable case (\ref{gammaproj}). In terms of the transverse-longitudinal decomposition (\ref{decomposition}), this is 
\begin{equation}
\Gamma^{i}=
 c_{1} \Delta^{3}h_{i}
 + 2 c_{2} \Delta^{2}\partial_{i}h^{T}
 + \gamma \Delta^{2}\partial_{i}\partial_{j}h_{j}.
\end{equation}
We have the expansion of the term, 
\begin{equation}
\begin{split}
	&
	\frac{1}{4}\left(\dot{n}^{i}-\Gamma^{i}\right)\mathfrak{D}^{-1}_{ij}\left(\dot{n}^{j}-\Gamma^{j}\right)=
	\frac{1}{4}\dot{n}^{iT}\frac{1}{\Delta^{2}}\dot{n}^{iT}
	-\frac{1}{4} \rho \dot{n}^{L}\frac{1}{\Delta}\dot{n}^{L}
	-\frac{1}{2} c_{1} \dot{n}^{iT}\Delta h_i^{T}
	+ \rho c_{2} \dot{n}^{L}\Delta h^{T}
	\\ &
	+ \rho \nu \dot{n}^{L}\Delta h^{L}
	+ \frac{1}{4} c_{1}^{2} h_i^{T}\Delta^{4}h_i^{T}
	- \rho c_{2}^{2} h^{T}\Delta^{3}h^{T}
	- \rho c_{2} \nu h^{T} \Delta^{3} h^{L}
	- \frac{1}{4} \rho \nu^{2} h^{L} \Delta^{3}h^{L},
\end{split}
\label{gaugeS}
\end{equation}
with $\nu =2c_{1}+2c_{2}+c_{3}$. As in the projectable case, the terms with a odd time derivative in (\ref{BB}) and (\ref{gaugeS}) can be canceled by an appropriate setting of the constants $c_{1,2,3}$, which coincides with (\ref{settingconstants}) since the notation on these constants is the same. With this choice, the final path integral in the Lagrangian formalism of the $3+1$ nonprojectable Ho\v{r}ava theory, with the $z=3$ potential, results
\begin{equation}
	\begin{split}
	Z =& 
	\int \mathcal{D}h_{ij}^{TT} \mathcal{D}h^T \mathcal{D}h_i^T \mathcal{D}h^L
	\mathcal{D}n \mathcal{D}n^{iT} \mathcal{D} n^L \mathcal{D}\bar{C}_i^T \mathcal{D} C^{iT} \mathcal{D}\bar{C}^L \mathcal{D}C^L \mathcal{D}\xi \mathcal{D}\bar{\varepsilon} \mathcal{D}\varepsilon
	\\ &
	\exp \left[ i 
		\int dt d^{3}x \left\lbrace \bar{C}_{k}^{T}\left(\partial_{t}^{2}+2\Delta^{3}\right)C^{kT}
		-\bar{C}^{L}\left[\partial_{t}^{2}\Delta+4(1-\lambda)(1+\kappa)\Delta^{4}\right]C^{L}\right. \right. 
	\\
		&\left.-\frac{1}{4}h_{ij}^{TT}\left(\partial_{t}^{2}+\beta_{3}\Delta^{3}\right)h^{TT}_{ij}-h^{T}\left[\frac{1-2\lambda}{8}\partial_{t}^{2}
		+\left(\frac{3\beta_{3}}{8}
		  +\beta_{4}
		    +\lambda^{2}( 1 + \kappa )\right)\Delta^{3}\right]h^{T}\right.
	\\
		&\left.
		+\frac{1}{8}h_i^{T}\left(\partial^{2}_{t}\Delta+2\Delta^{4}\right)h_i^{T}
		- \frac{1-\lambda}{4} h^{L}\left[ \partial^{2}_{t}
		+ 4 (1-\lambda) (1+\kappa)\Delta^{3}\right]h^{L}\right.\\
		&\left.
		- \frac{1}{4} n^{iT} \left(\frac{\partial_{t}^{2}}{\Delta^{2}} 
		+ 2 \Delta\right) n^{iT}
		+ \frac{1}{4} \rho n^{L}\left(\frac{\partial_{t}^{2}}{\Delta}
		+ 4 (1-\lambda) (1+\kappa) \Delta^{2}\right)n^{L}
		\right.\\
		&\left.
		+ \frac{\lambda}{2} h^{T}\left(\partial_{t}^{2}
		+ 4 (1-\lambda)(1+\kappa)\Delta^{3}\right)h^{L}
		-\alpha_{3}n\Delta^{3}h^{T}-\alpha_{4}n\Delta^{3}n \right.\\
		&\left. \left.	
		+\alpha_{3}\xi\Delta^{3}h^{T}+2\alpha_{4}\xi\Delta^{3}n-2\alpha_{4}\bar{\varepsilon}\Delta^{3}\varepsilon \right\rbrace \right] \,.
	\end{split}
\label{finalquantumnonproj}
\end{equation}
In the set of propagators derived from this action, shown in appendix B, almost all of them are regular. The nonregular ones arise when the variables associated to the second-class constraints, $\xi$ and $\bar{\varepsilon},\varepsilon$, are involved. This confirms that the nonlocal Lagrangian (\ref{finalquantumnonproj}) leads to regular propagators for the original field variables, including the ghosts associated to the gauge fixing \cite{Barvinsky:2015kil}, but the presence of nonregular propagators persists, associated to the fact that the theory has second-class constraints, unlike the projectable case.


\section{Comparison with General Relativity}
As it is well known, the classical canonical action of general relativity written in ADM variables is
\begin{equation}
	S=
	\int dt d^{3}x
	\Big( \dot{g}_{ij}\pi^{ij} - N\mathcal{H} - N^{k}\mathcal{H}_k \Big) \,.
\end{equation}
The constraints are given by
\begin{eqnarray}
	\mathcal{H}_i &=& 
	-2 \nabla^k\pi_{ki} \,,
\\
	\mathcal{H}&=&
	\frac{1}{\sqrt{g}} \left( \pi^{ij}\pi_{ij} - \frac{1}{2} \pi^2 \right) -\sqrt{g}R \,,
\end{eqnarray}
and both constraints are of first class. $N$ and $N^i$ play the role of Lagrange multipliers. We denote them collectively by $\mathcal{H}_a = (\mathcal{H},\mathcal{H}_i)$, and $N^a = (N, N^i)$. 

For the BFV quantization \cite{Fradkin:1975sj} we introduce the canonical pair $(N^{a},\pi_a)$, hence we have the functions $G_A=(\mathcal{H}_a,\pi_a)$. For each of these functions we define the pair of fermionic ghosts $(\eta_A \,, \mathcal{P}^A )$, which can be split as $(\eta_1^{a},\mathcal{P}^{1}_a)$, $(\eta_2^{a},\mathcal{P}^{2}_a)$. The involution relations $\{G_A \,, G_B\} = U_{AB}^{C} G_C$ lead to the algebra of spacetime diffeomorphisms. There is an essential qualitative difference with the Ho\v{r}ava theory, since in general relativity the coefficients $U_{ab}^c$ depend on the canonical fields. This fact has important consequences in the BFV quantization \cite{Fradkin:1975cq,Fradkin:1975sj}. The gauge-fixed BFV path integral takes the form 
\begin{equation}
\begin{split} 
	Z = 
	\int \mathcal{D} g_{ij} \mathcal{D} \pi^{ij} \mathcal{D}N^a \mathcal{D}\pi_a 
	\mathcal{D}\eta^A \mathcal{D}\mathcal{P}_A 
	\exp \left[
	i \int dt d^{3}x
	\Big(\dot{g}_{ij}\pi^{ij}+\pi_a\dot{N}^{a}+\mathcal{P}_A\dot{\eta}^{A}-\mathcal{H}_\Psi\Big) \right] \,.
\end{split}
\end{equation}
In the $3+1$-dimensional spacetime the two canonical pairs $(g_{ij},\pi^{ij})$, $(N^a,\pi^a)$ sum $20$ degrees of freedom. The ghosts $(\eta_1^{a},\mathcal{P}^{1}_a)$, $(\eta_2^{a},\mathcal{P}^{2}_a)$ sum $16$ degrees. The substraction yields the usual four physical degrees of freedom in the phase space of quantum general relativity. The BRST charge takes the form
\begin{equation}
	\Omega=
	G_A\eta^{A}=\mathcal{H}_a\eta_1^{a}+\pi_a\eta_2^{a}
	-\frac{1}{2}U_{ab}^{c}\eta_1^{a}\eta_1^{b}\mathcal{P}^{1}_c \,.
\end{equation}
The gauge-fixed quantum Hamiltonian is defined by Eq.~(\ref{generalgfhamiltonian}), with $\mathcal{H}_1=0$. The appropriate form of the gauge-fixing fermionic function is given in (\ref{Psi}), which, considering the four spacetime directions, takes the form $\Psi = \mathcal{P}_a^{1}N^{a}+\mathcal{P}_a^{2}\chi^{a}$.
Thus, the gauge-fixed Hamiltonian results
\begin{equation}
	\begin{split}
	\mathcal{H}_\Psi =&
	N^{a}\mathcal{H}_a+\mathcal{P}_a^{1}\eta_2^{a}+\pi_a\chi^{a}+\mathcal{P}_a^{2}\{\chi^{a} \,, \mathcal{H}_b\} \eta_1^{b}
	+\mathcal{P}_a^{2}\{\chi^{a} \,, \pi_b\} \eta_2^{b}
	\\
	& 
	-\frac{1}{2}\mathcal{P}_a^{2}\{\chi^{a},U_{cd}^{e}\} \eta_1^c \eta_1^d \mathcal{P}_e^1
	+ N^{a} \eta_1^{b} U_{ba}^c \mathcal{P}_c^{1} \,.
	\end{split}
\end{equation}

We proceed to the construction of the quantum Lagrangian. For the integration on $\pi_a$ we adopt the same strategy we used in the Ho\v{r}ava theory, considering in this case the four directions of spacetime diffeomorphisms. We take a gauge-fixing condition in the form 
\begin{equation}
 \chi^a = 
 \mathfrak{D}^{ab}\pi_b + \Gamma^a[g_{ij}, N, N^i] \,.
\end{equation}
The isotropic scaling in general relativity is 
\begin{equation}
	[\pi_a]=3,\qquad [\chi^{a}]=1,\qquad [\mathfrak{D}^{ab}]=-2 \,.
\end{equation}
Therefore, $\mathfrak{D}^{ab}$ is nonlocal whereas its inverse $\mathfrak{D}_{ab}^{-1}$ is a local operator. After the integration on $\pi_a$, the quantum action takes the form 
\begin{equation}
	\begin{split}
	S=&
	\int\,dt\,d^{3}x\Big(
	\dot{g}_{ij}\pi^{ij}
	+ \frac{1}{2} (\dot{N}^{a} - \Gamma^{a}) \mathfrak{D}^{-1}_{ab} 	(\dot{N}^{b} - \Gamma^{b})
	+ \mathcal{P}_A\dot{\eta}^{A}
	- N^{a}\mathcal{H}_a
	- \mathcal{P}_a^{1}\eta_2^{a}
	\\ &
	- N^{a} \eta_1^{b} U_{ba}^c \mathcal{P}_c^{1}
	- \mathcal{P}_a^{2} \{\Gamma^{a} \,, \mathcal{H}_b \}\eta_1^{b}
	- \mathcal{P}_a^{2} \dfrac{\delta \Gamma^{a}}{\delta N^b} \eta_2^{b}
	+ \frac{1}{2} \mathcal{P}_a^{2} \{\Gamma^{a},U_{bc}^d \} \eta_1^b\eta_1^c\mathcal{P}_d^{1}
	\Big)\,.
	\end{split}
\end{equation}
The ghost sector is given by
\begin{equation}
	-\bar{\mathcal{P}}_a\mathcal{P}^{a}
	+ \bar{\mathcal{P}}_a \left( \dot{C}^{a} - \frac{1}{2} \bar{C}_d \{\Gamma^{d} \,, U_{bc}^{a}\} C^b C^c 
	+ N^{d}C^{b}U_{bd}^{a} \right)
	+ \mathcal{P}^{a} \left( \dot{\bar{C}}_{a}+\bar{C}_e \dfrac{\delta \Gamma^e}{\delta N^a} \right) \,.
\end{equation}
By integrating on the corresponding Grassmann variables we get the action
\begin{equation}
	\begin{split}
	S=&
	\int\,dt\,d^{3}x\Big(
	\dot{g}_{ij}\pi^{ij}
	+\frac{1}{2} (\dot{N}^{a} - \Gamma^{a}) \mathfrak{D}^{-1}_{ab} (\dot{N}^{b} - \Gamma^{b})
	- N^{a}\mathcal{H}_a  \\
	&
	+ \left( \dot{C}^{a}
	-\frac{1}{2}\bar{C}_d\{\Gamma^{d},U_{bc}^{a}\}C^{b}C^{c}
	+N^{d}C^{b}U_{bd}^{a}\right)
	\left( \dot{\bar{C}}_{a}+\bar{C}_e \dfrac{\delta \Gamma^e}{\delta N^a} \right)
	-\bar{C}_a\{\Gamma^{a},\mathcal{H}_b \}C^{b} \Big) \,.
	\end{split}
\end{equation}

We now perform perturbations, obtaining the second-order action
\begin{equation}
	\begin{split}
	S=&
	\int\,dt\,d^{3}x \left[
	-\left( p^{TT}_{ij} - \frac{1}{2}\dot{h}_{ij}^{TT} \right)^2
	+\frac{1}{4}\dot{h}^{TT}_{ij}\dot{h}^{TT}_{ij}
    +\frac{1}{2}p^{T}\dot{h}^{T}
    + \frac{1}{2}p^k \Delta p^k \right.
    \\ &
	+p^{k}(-\partial_kp^T+(\delta_{kj}\Delta
	+\partial_k\partial_j)(n^j-\frac{1}{2}\dot{h}_j))
	+\frac{1}{4}h^{TT}_{ij}\partial^{2}h^{TT}_{ij}
	-\frac{1}{8}h^{T}\partial^{2}h^{T}
	-n\partial^{2}h^{T}
	\\ & \left.
	+\frac{1}{2} (\dot{N}^{a} - \Gamma^{a}) \mathfrak{D}^{-1}_{ab} (\dot{N}^{b} - \Gamma^{b})
    -\bar{C}_a\{\Gamma^{a},\mathcal{H}_b \}C^{b}
	+ \dot{C}^{a}
	\left( \dot{\bar{C}}_{a}+\bar{C}_e \dfrac{\delta \Gamma^e}{\delta N^a} \right) \right] \,.
	\end{split}
\label{grprepiij}
\end{equation}
Here we face another qualitative difference with respect to the Ho\v{r}ava gravity. The Lagrangian in (\ref{grprepiij}) has no $(p^T)^2$ term, unlike the Lagrangian in Eq.~(\ref{actionprepijpertur}). This is a consequence of the relativistic structure behind the Hamiltonian of general relativity, which implies the frozen of the scalar mode. Hence, we change the order of integration in this case, by performing first the integration on the longitudinal component of the momentum $p^i$. This brings the terms to the Lagrangian
\begin{equation}
	\begin{split}
	&
	+ \frac{1}{2} (p^T)^2 
	+ p^T(-2 \partial_kn^k + \partial_k\dot{h}_k)
	- \frac{1}{2} n^k(\delta_{kl}\Delta + 3\partial_k\partial_l)n^l \\
	&
	- \frac{1}{8}\dot{h}_k(\delta_{kl}\Delta +3\partial_k\partial_l)\dot{h}_l
	+ \frac{1}{2} n^k(\delta_{kl}\Delta 
	   + 3\partial_k\partial_l)\dot{h}_l \,.
	\end{split}
\end{equation}
Now we perform the integration on $p^T$ and $p^{ij}_{TT}$, obtaining
\begin{equation}
	\begin{split}
	S=&
	\int dt d^{3}x\Big(
	\frac{1}{4}\dot{h}_{ij}^{TT}\dot{h}_{ij}^{TT}
	-\frac{1}{8}\dot{h}_k^T \Delta \dot{h}_k^T
	+\frac{1}{4}h^{TT}_{ij}\Delta h^{TT}_{ij}
	-\frac{1}{8}h^{T}\Delta h^{T}
	-n\Delta h^{T}
	-\frac{1}{2}n^{kT}\Delta n^{kT}
	\\
	&
	+\frac{1}{2}n^{kT} \Delta \dot{h}^T_k
	-\frac{1}{8}\Big( (\dot{h}^T)^2 + \dot{h}^T(-8\Delta n^L
	+4\dot{h}^L)\Big)-\bar{C}_a\{\Gamma^{a},\mathcal{H}_i \} C^{i}
	\\&
	+\dot{C}^{a}(\dot{\bar{C}}_{a}+\bar{C}_e \dfrac{\delta \Gamma^e}{\delta N^a})
	+\frac{1}{2} (\dot{n}^{a}-\Gamma^{a}) \mathfrak{D}_{ab}^{-1} (\dot{n}^{b}-\Gamma^{b})\Big) \,.
	\end{split}
\end{equation}
Therefore, the resulting quantum Lagrangian is completely local as far as the remaining part $\Gamma^a$ of the gauge-fixing condition is local.

\section*{Conclusions}
We have seen that the BFV quantization is suitable for the Ho\v{r}ava theory, both in its projectable and nonprojectable versions, and varying the dimension of the foliation. This extends the analysis that two of us performed in Ref.~\cite{Bellorin:2021tkk}. The BFV formalism provides a rich framework to study the quantum dynamics of the Ho\v{r}ava gravity, in particular by incorporating the BRST symmetry in terms of the canonical variables. In the past it has been used to establish the unitarity of gauge theories, thanks to the ability of introducing a bigger class of gauge-fixing conditions in the Hamiltonian formalism \cite{Fradkin:1975cq,Fradkin:1975sj}.

We have seen that the BFV version of the projectable (three-dimensional) theory reproduces the quantum Lagrangian presented in Ref.~\cite{Barvinsky:2015kil}, which was obtained by fixing the gauge following the Faddeev-Popov procedure. Our results reinforces the consistency of the quantization of the theory. We have performed the integration on momenta after specifying the dependence that the gauge-fixing condition has on them. Specifically, we have introduced a linear dependence on the momentum conjugated to the shift vector. Guided by a criterium of anisotropic scaling, we have incorporated an operator that balances the momentum in the gauge-fixing condition. It turns out that, in both versions of the Ho\v{r}ava theory, this operator introduces a nonlocality in the Lagrangian after the integration. Thus, we have arrived at the same result obtained in \cite{Barvinsky:2015kil} of having a nonlocal quantum Lagrangian, in our case starting from a self-consistent Hamiltonian formulation provided by the BFV formalism. The original Hamiltonian theory is completely local. In Ref.~\cite{Barvinsky:2015kil} it was pointed out that the final nonlocality of the quantum Lagrangian, restricted to the kinetic term of the shift vector, can be eliminated by introducing the conjugated momentum of the shift vector. We have corroborated this in an inverse way, starting from the complete, self-consistent and local Hamiltonian formulation and ending with the nonlocal Lagrangian. With the aim of having a further comparison, we have performed the same procedure in general relativity, taking into account the relativistic isotropy of its field variables. In this case the operator introduced in the gauge fixing-condition is nonlocal and the quantum Lagrangian resulting after the integration is local (whenever the dependence of the gauge-fixing condition on the rest of variables is local). Thus, we see an interesting relationship between the anisotropy of the underlying symmetry and the nonlocality of the quantum Lagrangian. The relationship has been established on very basic grounds, since it comes from the integration of the Hamiltonian theory.

\section*{Acknowledgments}

C.~B.~is partially supported by the grant CONICYT PFCHA/DOCTORADO BECAS CHILE/2019 -- 21190960. C.~B.~is a graduate student in the ``Doctorado en F\'isica Menci\'on F\'isica-Matem\'atica" PhD program at the Universidad de Antofagasta. B.~D.~is partially supported by the Universidad de Antofagasta grant PROYECTO ANT1756, Chile.


\appendix
\section{Nonperturbative integration on $\pi^{ij}$}
\label{integratepi}
All the dependence that the action given in (\ref{prepiij}) has on $\pi^{ij}$ is contained in the three terms
\begin{equation}
	\pi^{ij}\dot{g}_{ij}-\mathcal{H}-N^{k}\mathcal{H}_k =
	- \frac{1}{\sqrt{g}} \mathcal{G}_{ijkl} \pi^{ij} \pi^{kl} + 2 K_{kl}\pi^{kl}
	- \sqrt{g}\mathcal{V} \,,
	\label{termspiij}
\end{equation}
where
\begin{eqnarray}
	\mathcal{G}_{ijkl}&=&
	\frac{1}{2} ( g_{ik}g_{jl} + g_{il}g_{jk} ) 
	- \frac{\lambda}{1-d\lambda} g_{ij}g_{kl} \,,\\
	{K}_{kl} &=& 
	\frac{1}{2} \left( \dot{g}_{kl} - 2 \nabla_{(k} N_{l)} \right) \,.
\end{eqnarray}
After completing the square involving $\pi^{ij}$ and making the Gaussian integration, from the terms in (\ref{termspiij}) there results
\begin{equation}
	\sqrt{g} G^{ijkl} K_{ij} K_{kl} - \sqrt{g} \mathcal{V}
	= \mathcal{L}_{\mbox{\tiny cl}} \,,
\end{equation}
where ${G}^{ijkl} = \frac{1}{2} ( g^{ik}g^{jl} + g^{il}g^{jk} ) 
- \lambda g^{ij}g^{kl}$, such that it is the inverse of $\mathcal{G}_{ijkl}$ \cite{Horava:2009uw}, and $\mathcal{L}_{\mbox{\tiny cl}}$ coincides with the classical Lagrangian (\ref{lagrangian}). A factor of $(\det \mathcal{G})^{-1/2}$ arises in the measure after the integration. Therefore, the path integral takes the form
\begin{equation}
	\begin{split}
		Z =&
		\int \mathcal{D} g_{ij} \mathcal{D} N^{k}\mathcal{D}\bar{C}_i \mathcal{D} C^i (\det \mathcal{G})^{-1/2}
		\exp\left[ i \int dt d^{d}x 
		\Big( 
		\mathcal{L}_{\mbox{\tiny cl}}
		+ \frac{1}{4} (\dot{N}^{k}-\Gamma^{k}) \mathfrak{D}^{-1}_{kl} (\dot{N}^{l}-\Gamma^{l})
		\right.  \\
		& \left.		 
		+\left(\dot{C}^{k}-N^{j}\partial_{j}C^{k}-N^{k}\partial_{j}C^{j}\right)\left(\dot{\bar{C}}_k+\bar{C}_i\frac{\delta\Gamma^{i}}{\delta N^{k}}\right)				
		- \bar{C}_i \{ \Gamma^{i} \,, \mathcal{H}_k \} C^{k} 
		\Big) \right] \,.
	\end{split}
\end{equation}


\section{Propagators of the nonprojectable theory}
By making Fourier transforms in time and space in the Lagrangian (\ref{finalquantumnonproj}), and inverting the resulting operator between the squared fields, we obtain the propagators of the nonprojectable theory,
\begin{align}
 &
 \langle h_{ij}^{TT}h_{ij}^{TT}\rangle= P_1 \,,
 \qquad  
 \langle h^{T}h^{T}\rangle= \frac{2(1-\lambda)}{(1-3\lambda)} P_2  \,,
 \qquad 
 k^2 \langle h_i^{T}h_i^{T}\rangle= 2 P_3 \,,
 \nonumber \\ &
 \langle h^{L}h^{L}\rangle=
 \frac{1}{1-\lambda} P_4 
 + \frac{ 2 \lambda^{2}}{(1-\lambda)(1-3\lambda)} P_2  \,,
 \qquad 
 \langle h^{T}h^{L}\rangle=
 \frac{4 \lambda}{(1-3\lambda)} P_2 \,,
 \nonumber \\ &
 \langle nn\rangle= 
 \frac{\alpha_{3}^{2}(1-\lambda)}{2\alpha_{4}(1-3\lambda)} P_2 \,,
 \qquad
 \langle nh^{T}\rangle=
 -\frac{2\alpha_{3}(1-\lambda)}{\alpha_{4}(1-3\lambda)} P_2 \,,
 \nonumber \\ &
 \langle nh^{L}\rangle=
 -\frac{2\alpha_{3}\lambda}{\alpha_{4}(1-3\lambda)} P_2 \,,
 \qquad
 \langle n^{iT}n^{iT}\rangle= k^{4} P_3 \,,
 \qquad
 k^2 \langle n^{L}n^{L}\rangle= (1+\kappa) k^4  P_4 \,,
 \nonumber \\ &
 \langle\bar{C}_{k}^{T}C^{kT}\rangle= P_3 \,,
 \qquad
 k^2 \langle\bar{C}^{L}C^{L}\rangle= P_4 \,,
 \qquad
 \langle n\xi\rangle= - 2 P_5 \,,
 \nonumber \\ &
 \langle\xi\xi\rangle= - P_5 \,,
 \qquad
 \langle \bar{\varepsilon}\varepsilon\rangle= - 2 P_5 \,,
 \nonumber \\ &
 \label{propagtaors}
\end{align}
where
\begin{align}
	&
	P_1 = \frac{4}{\omega^{2}+\beta_{3}k^{6}} \,, 
	\nonumber \\ &
	P_2 = 4 \left[ \omega^{2}
	+\frac{(1-\lambda)}{\alpha_{4}(1-3\lambda)}
	\left(\alpha_{4} \left( 3\beta_{3} + 8 \beta_{4} \right) - 2 \alpha_{3}^{2}\right) k^{6} \right]^{-1} \,,
	\nonumber \\ &
	P_3 = \frac{4}{\omega^{2}+2k^{6}} \,,
	\qquad
	P_4 = \frac{4}{\omega^{2}+4(1-\lambda)(1+\kappa)k^{6}} \,,
	\nonumber \\ &
	P_5 = \frac{1}{\alpha_{4}k^{6}} \,.
	\nonumber \\ &
\end{align}
In (\ref{propagtaors}) we have written the propagators of $h_i^T$, $n^L$, and $\bar{C}^L,C^L$ multiplied by $k$ factors since these variables must be compensated with spatial derivatives in the composition of the original tensor variables. We adopt the definition of regular propagators given in Ref.~\cite{Barvinsky:2015kil}, which is based on the analysis of renormalizability of Lorentz-violating gauge theories of Refs.~\cite{Anselmi:2007ri,Anselmi:2008bq,Anselmi:2008bs}. $P_1,P_2,P_3$, and $P_4$ are regular, but $P_5$ is not regular.\footnote{Actually, to hold the regularity of $P_1,P_2,P_3$, and $P_4$ several conditions on the constants must be imposed, since the coefficients of $\omega$ and $k^6$ must be positive (after a Wick rotation) and unwanted ghosts should be avoided. The number of independent constants is enough to have nonempty sets where these conditions holds.} Thus, the nonregularity is located on the variables $\xi,\bar{\varepsilon},\varepsilon$ associated to the second-class constraints. 



\end{document}